\documentclass[acmlarge]{acmart}
\makeatletter
\newcommand{\confshort}{\acmConference@shortname}
\newcommand{\conffull}{\acmConference@name}
\newcommand{\confdate}{\acmConference@date}
\newcommand{\confloc}{\acmConference@venue}
\AtBeginDocument{
  \fancypagestyle{firstpagestyle}{
    \fancyhead{}%
    \fancyfoot[C]{}%
  }
  \fancyhf{}
  \fancyhead[LO]{\@headfootfont\shorttitle}%
  \fancyhead[RE]{\@headfootfont\@shortauthors}%
  \fancyhead[LE]{\@headfootfont\footnotesize \confshort, \confdate, \confloc}%
  \fancyhead[RO]{\@headfootfont\footnotesize \confshort, \confdate, \confloc}%
  \fancyfoot[C]{}%
}
\makeatother
\acmBooktitle{\conffull\@ (\confshort), \confdate, \confloc}

\usepackage{tabularx}
\usepackage{array}
\usepackage[inline]{enumitem} % inline enumeration
\usepackage{xcolor} % highlight table
\usepackage{longtable}
\usepackage{booktabs}
\usepackage{todonotes}

\AtBeginDocument{%
  }

\copyrightyear{2026}
\acmYear{2026}
\setcopyright{cc}
\setcctype{by-nc-nd}
\acmConference[FAccT '26]{The 2026 ACM Conference on Fairness, Accountability, and Transparency}{June 25--28, 2026}{Montreal, QC, Canada}
\acmBooktitle{The 2026 ACM Conference on Fairness, Accountability, and Transparency (FAccT '26), June 25--28, 2026, Montreal, QC, Canada}
\acmDOI{10.1145/3805689.3812208}
\acmISBN{979-8-4007-2596-8/2026/06}

\begin{document}

\title[Do Language Models Pass the Bechdel Test?]{Do Language Models Pass the Bechdel Test? Auditing Gender Biases in LLM-Generated Screenplays}

\author{Megha N. Govindu}
\email{meggov@sas.upenn.edu}
\orcid{0009-0005-3232-2332}
\affiliation{%
  \institution{University of Pennsylvania}
  \city{Philadelphia}
  \state{Pennsylvania}
  \country{USA}
}

\author{Stephanie T. Wang}
\email{stephtw@seas.upenn.edu}
\orcid{0009-0002-3437-2227}
\affiliation{%
  \institution{University of Pennsylvania}
  \city{Philadelphia}
  \state{Pennsylvania}
  \country{USA}
}

\author{Sorelle A. Friedler}
\email{sorelle@cs.haverford.edu}
\orcid{0000-0001-6023-1597}
\affiliation{%
  \institution{Haverford College}
  \city{Haverford}
  \state{Pennsylvania}
  \country{USA}
}

\author{Danaé Metaxa}
\email{metaxa@upenn.edu}
\orcid{0000-0001-9359-6090}
\affiliation{%
  \institution{University of Pennsylvania}
  \city{Philadelphia}
  \state{Pennsylvania}
  \country{USA}
}

\renewcommand{\shortauthors}{Govindu et al.}

\begin{abstract}
As large language models (LLMs) are increasingly used in media production from journalism to filmmaking, what impact do they have on the stories being told? Prior work has shown LLMs to perpetuate social biases, including those related to gender. We complement existing literature on gender bias in LLM outputs by auditing the network structure of LLM-generated movie screenplays through automating the Bechdel test, a popular measure of women’s representation in literary and film works. We also introduce the use of social network analysis measures to further analyze representational bias in LLM-generated scripts. We evaluate screenplays generated by three state-of-the-art LLMs (GPT-5, Gemini 3 Pro, and Claude Sonnet 4.5) against 768 corresponding human-written screenplays, finding that human-written scripts are more likely to pass the Bechdel test. However, other network analyses, like centrality, homophily, and triadic relationships demonstrate that in some cases LLM-scripts have less bias, although all script types demonstrate some representational bias under most measures. We conclude by discussing the continued need for further quantitative assessments of media representations and AI-generated content.
\end{abstract}



\received{13 January 2026}
\received[revised]{25 March 2026}
\received[accepted]{16 April 2026}

\maketitle

\section{Introduction}

In 2023, The Writers Guild of America (WGA) went on a 148 day strike, with AI usage in screenwriting a primary aspect in negotiations with the Alliance of Motion Picture and Television Producers~\cite{kinder_hollywood_nodate}. Screenwriters on strike warned that AI could not provide a complex representation of human life, and especially of the experiences of marginalized groups: ``They want to replace us with AI but I can tell you now definitively that AI does not have the trauma, the joy, of the lived experience~\cite{glaad2023}.'' Since the strike, the move towards replacing screenwriters has continued. In 2024, Lionsgate gave a startup permission to train their new model on the Lionsgate content library~\cite{lee_lionsgate_2024}. Most recently, OpenAI is backing an animated feature-length film made with an ``AI-assisted workflow,'' set to debut at the Cannes Film Festival in 2026~\cite{toonkel_openai_2025}.

We learn about ourselves and the world at the movies. Representation of marginalized identities in film shapes self-identity, experiences of discrimination, and individuals' capacities to imagine roles in society for themselves and others~\cite{saleem_muslim_2019, leavitt_frozen_2015}.  Yet there is copious, well-documented evidence that large language models (LLMs) and other forms of generative AI exhibit biases on the basis of race~\cite{hofmann_ai_2024}, gender~\cite{wan_kelly_2023, lucy_gender_2021}, and a range of other attributes, as well as various intersections (see Section \ref{sec:related_work_llm_bias}). Given this context, we ask: as LLMs are increasingly used in the film industry, will such biases affect the artistic content being produced?

This paper takes up this question by examining potential gender bias against women in LLM-generated movie scripts, and using these screenplays as a lens through which to examine LLMs for representational bias. One popular measure of the representation of women in film is the Bechdel test. A film passes the test if there are 1) two women who 2) talk to each other about 3) something besides a man \cite{bechdel_dykes_1985}. We implement the Bechdel test and use it to measure the representation of women in LLM-generated scripts from three state-of-the-art LLMs (GPT-5, Gemini 3 Pro, and Claude Sonnet 4.5) and 768 corresponding real, human-written movie screenplays. Screenplays are generated using synopses of films whose real scripts are available and prompting the LLMs to first create scene lists and then scene-by-scene to generate scripts for those scenes. To analyze this data, our implementation of the Bechdel test creates a character network where each character is a node and each dialog interaction in the script is represented as an edge in the network. Using this network, we further analyze the representation of women characters on a range of social network analysis (SNA) measures, including homophily and centrality.

Our results show that human-written scripts are far more likely to pass the Bechdel test than LLM-generated ones. Examining other representational bias measures based on character networks yields more varied results; GPT-generated scripts outperform human ones on proportion of interactions involving women, and also have more heterophily and triadic relationships between female characters. 

While these relative representational differences between generated and human scripts are important, overall we find that none of these script types have stronger representation of women than men; across all script types and centrality measures studied, men are more central within the network than women. The Bechdel test was introduced to point out the lack of representation of women in media; our representational analysis shows that these concerns remain across both human and LLM-generated screenplays.

\section{Related Work}
This study builds upon three main areas of work: media studies, analyses of social network representations of interpersonal relationships, and LLM bias audits.

\subsection{Biases in Film and the Bechdel Test}
% Why is studying representation in film important?
Considerable research across disciplines from communication to media studies has examined gender and racial/ethnic representation in film and literature. Such work is motivated by the understanding that film, as a dominant form of media culture, both reflects and shapes social values and identities, offering insight into the contexts in which cultural narratives are produced~\cite{kellner_media_1995}.  

% Ways prior work has studied representation in film and other media
Methodologically, prior work relied on highly manual studies in which researchers code the gender and ethnicity of characters to examine their presence and centrality within narratives~\cite{smith_inequality_2023, mccabe_gender_2011, tukachinsky_documenting_2015}. More recently, advances in computational methods and data availability have enabled large-scale, data-driven approaches to quantifying stereotypes and biases in film~\cite{kumar_gender_2022, bamman_measuring_2024, sap_connotation_2017, yu_unpacking_2022}. For instance, \citet{bamman_measuring_2024} use computer vision to analyze gender and race representation across 2,307 films, revealing increasing diversity over time but persistent under-representation of Black actors in award-nominated films and of non-White actors and women in leading roles. Other works analyze screenplays for gender bias using connotation frames~\cite{sap_connotation_2017} and by analyzing socio- and psycho-linguistic dimensions of dialogue~\cite{yu_unpacking_2022}.

% Introduce the Bechdel test
In 1985, Alison Bechdel's comic strip ``The Rule'' introduced what became known as the Bechdel test. In order to pass the test, a given film or piece of fiction must satisfy three conditions: it must (1) have at least two named women in it, (2) who talk to each other, (3) about something other than a man~\cite{bechdel_dykes_1985}. Originally a humorous commentary and critique, the Bechdel test has since become a mainstream heuristic for benchmarking female representation in media \cite{Valentowitsch_2023}.

% Bechdel test in research
Research has adapted the Bechdel test for computational analysis of gender representation at scale. \citet{garcia_gender_2014} quantify the Bechdel test by constructing a Bechdel score and applying this to movie scripts and MySpace and Twitter dialogue datasets to compare gender roles in both fictional and real-world online dialogues. \citet{agarwal_key_2015} builds on this work to automate the Bechdel test, building a classifier and studying the effectiveness of various linguistic and SNA features, finding that SNA features are more effective at accurately labeling scripts with Bechdel scores. They also find that female characters exhibit lower centrality (in the movie's social network) in movies that fail the Bechdel test.
Building on~\citet{agarwal_key_2015}, we implement a computational Bechdel test and validate our implementation against theirs before extending to it to machine-generated screenplays. We apply this test alongside other network measures to to both human-written and LLM-generated screenplays to evaluate the degree to which gender biases in screenplays are amplified by language models.

\subsection{Social Network Analysis} 
% Social Network Analysis.... Origins?
% Social network analysis (SNA) enables us to model complex social structures in a concrete network representation, where nodes represent entities (i.e. people) while edges represent relationships between them (i.e. interactions between people). 
% ~\cite{barnes_edge_2025} 
Social network analysis (SNA) uses network structures to model social structures, revealing patterns in group formation, influence, and connectivity. In SNA, networks are built of nodes (representing entities, like people) and edges (relationships between those entities). These methods have been applied to study human relationships in online social networks~\cite{ugander_anatomy_2011} and, more recently, to understand structural bias and demographic-driven disparities in social networks. For a survey, see~\citet{saxena_fairsna_2024}.

% Movie analysis with social network analysis
Social network analysis has also been used to analyze media such as film, producing empirical analyses of social ties among film characters~\cite{kumar_gender_2022, agarwal_key_2015, park_character-net_2009, lv_storyrolenet_2018, kagan_using_2020, weng_rolenet_2007}. \citet{agarwal_parsing_2014} develop methods for parsing movie screenplays into social networks to study character interactions. \citet{weng_rolenet_2007} introduce RoleNet, a method that constructs social networks by connecting characters appearing in the same scene, and develops an algorithm to classify characters into leading or supporting roles and identify various structures of social ties. \citet{park_character-net_2009} build upon these ideas with Character-Net, which constructs the network from the accumulation of dialogue graphs between characters, treating dialogue as a relationship between two people. They use Character-Net to develop an algorithm that classifies characters into major, minor, and extra roles. StoryRoleNet is another project that integrates both video-based and subtitle-based networks by segmenting films into story units and computing relationship weights from both modalities within each unit before merging them into a social network for analysis~\cite{lv_storyrolenet_2018}. \citet{kagan_using_2020} also construct a social network representation of movies, and use SNA measures such as centrality and triangle analysis to analyze gender biases in women's portrayal in film. They also develop an automated classifier to predict whether a film passes the Bechdel test. 

% \subsection{LLMs for Social Network Generation}
The methods above analyze social networks extracted from existing media. Recent work has begun applying these methods to LLM-generated networks as LLMs become capable of social simulation~\cite{park_generative_2023}. 
Recent work related to SNA explores LLM's abilities to \textit{generate} social networks, and to simulate interactions over those networks. \citet{papachristou_network_2025} study the network formation behaviors of multiple LLM agents, benchmarking them against human decisions. They find that LLMs exhibit differences in micro- and macro-level properties across different social contexts (e.g., homophily in friendship networks, heterophily in organizational settings) in ways similar to human social networks. \citet{chang_llms_2025} explore how realistic social networks generated by LLMs compare to real ones, and examine demographic biases in social ties. They find differences in realistic-ness of generated networks among three prompting methods, and find that LLMs overestimate political homophily compared to real networks.  

Instead of explicitly prompting LLMs to generate networks, we prompt them to generate film screenplays, then parse and extract social networks from the resulting output following~\citet{agarwal_key_2015, agarwal_parsing_2014}. We validate the quality of the extracted networks by confirming the computational Bechdel test performs comparably on them to human-written ones, then apply it alongside character role classification~\cite{park_character-net_2009} and SNA measures --- including homophily, triangle analysis and centrality --- to characterize the position of female characters within the implicit networks that emerge from LLM-generated narratives.

% evidence of racial steering, default whiteness, and steering of minority homeseekers toward neighborhoods with lower opportu-nity indices in GPT-4’s housing recommendations to prospective buyers or renters~\cite{liu_racial_2024}.
\subsection{Audits of LLM Biases} 
\label{sec:related_work_llm_bias}
Algorithm auditing is a technique for systematically probing black-box systems through repeated interaction to understand their behavior and investigate potential biases without direct access to their internals~\cite{metaxa_auditing_2021}. Such methods have been commonly used to study biases by attributes like gender or political leaning, in domains including search engines~\cite{metaxa_search_2019, robertson_auditing_2018}, social media~\cite{wang_lower_2024}, and targeted advertising~\cite{lambrecht_algorithmic_2019}. 
We apply this auditing framework to large language models, systematically probing their screenplay generation to uncover potential gender biases in narrative representation. 

Prior work has identified two major types of harms resulting from bias in LLM models, termed \textit{allocative} and \textit{representational} ~\cite{blodgett_language_2020, barocas_problem_2017, crawford_trouble_2017}. 
Earlier work on word embeddings found that they embed human-like biases~\cite{caliskan_semantics_2017} and entrench gender stereotypes~\cite{bolukbasi_man_2016}, and proposed methodologies for modifying embeddings to address these bias sources. An ongoing line of research  documented numerous forms of bias in LLMs, including gender bias~\cite{lucy_gender_2021, wan_kelly_2023, wilson_gender_2024}, differential expressions of respect across genders, races and sexual orientation~\cite{sheng_woman_2019}, biases in occupational associations~\cite{kirk_bias_2021}, racial bias~\cite{hofmann_ai_2024, liu_racial_2024}, biases associating Muslims with violence~\cite{abid_persistent_2021}, and stereotyping ~\cite{siddique_who_2024}.

As LLMs are increasingly integrated into human studies research, work has found they can misportray and flatten the representation of marginalized demographic groups when used as proxies for human participants~\cite{wang_large_2025}. In a context related to movies, an audit of GPT’s content moderation system found that many real and generated TV scripts are flagged as content violations, with flagging strongly linked to age rating, genre, and violent content. The study brought to attention the censorship implications when LLM moderation systematically shapes the production of cultural narratives~\cite{mahomed_auditing_2024}, a question we take up from another angle in this work. 

Complementing existing bias studies of LLM systems, we conduct a social network–based audit of three state-of-the-art LLMs, prompting each to generate 768 film screenplays and examining whether and how these models exhibit gender bias in their narrative outputs.

\section{Dataset Design and Collection}
To compare gender bias in LLM outputs relative to human outputs, we compile a dataset of human-written scripts and their anonymized synopses. Then, we use these synopses to generate a dataset of counterpart LLM-generated movie scripts from three models: GPT 5, Gemini 3 Pro, and Claude Sonnet 4.5. 

\subsection{Human-written Screenplays} 
We collected 1,116 human-written, English-language screenplays from The Internet Movie Script Database\footnote{https://imsdb.com/}(IMSDb) and corresponding metadata --- including short synopsis, cast information, genre, and rating --- from The Movie Database\footnote{https://www.themoviedb.org/} (TMDB), both publicly available sources used in prior computational and auditing work on screen media~\cite{garcia_gender_2014, gorinski_movie_2015, agarwal_key_2015, proebsting_identity-related_2025}. Our selection was constrained to movie titles listed on IMSDb as of February 2025. Both sites are actively maintained; IMSDb requires administrator approval for script submissions, TMDB has a moderator system, and both include movies released as recently as 2025, the time of data collection. While IMSDb exclusively features screenplays that have been ``made into movies,'' other site-level curation criteria are unknown.

From the initial 1,116 screenplays, we applied two filtering criteria for data completeness. First, we retained only screenplays with corresponding Bechdel test scores on the Bechdel Test Movie List\footnote{https://bechdeltest.com/}, reducing the dataset to 825 movies. Second, we manually reviewed and excluded screenplays with synopses lacking meaningful plot details --- synopses that did not describe story events, characters, setting or conflicts specific to the film, but instead described only its production context (e.g., director, cast, soundtrack), technical properties (e.g., lighting, film format), relationship to other works (e.g., sequel, remake, reboot), or reception and framing (e.g., critical reception, filmmaker's intention). For example, we excluded the synopses ``A remake of the Wes Craven cannibal horror film'' (\textit{The Hills Have Eyes}, 2006) or ``A short work by Si Fried, screened by the Ann Arbor Film Festival in 1972'' (\textit{Out of Site}, 1971). This yielded a final dataset of 783 screenplays with Bechdel metadata and meaningful plot synopses.

The finalized dataset spans more than a century of cinema (see Appendix Table{~\ref{tab:films-full}} for the full list of films included in our dataset), from \textit{Jane Eyre} (1910) to \textit{Anora} (2024), with additional examples including \textit{Pulp Fiction} (1994), \textit{The Perks of Being a Wallflower} (2012), and \textit{Deadpool} (2016). None of the included films are known to have been primarily generated by AI. Screenplays averaged 23,672 words in length (SD = 4,840). Movies were tagged with multiple genres, such as `Thriller, Action' and `Drama, Romance, Comedy,' and indicated as unrated, G, PG, PG-13, R, or NC-17.

\subsection{Screenplay Generation} 

\subsubsection{Model selection}
We focus on three state-of-the-art LLMs, models underlying some of the most widely used chatbot systems at the time of writing~\cite{bailyn_top_2025}: GPT-5, Gemini 3 Pro, and Claude Sonnet 4.5.  
All experiments were conducted via the models' respective APIs in December 2025.

\subsubsection{Synopsis anonymization}
To generate full length screenplays, we prompted the LLMs using anonymized short synopses from TMDB for the 783 human-written screenplays in our dataset. Since we intended for the LLMs to generate the characters and their interactions, we cleaned all identifying information from the synopses -- apart from including the synopsis and scene count in the prompt, we did not include any other information about the movie (number of characters, genre, year, title etc.). Synopses were anonymized by replacing character names with generic identifiers (e.g., PERSON 1), substituting gendered relational terms with neutral equivalents (e.g., ``husband'' or ``wife'' with ``spouse''), and replacing gendered pronouns with gender-neutral ones. For example, \textit{Confessions of a Dangerous Mind} (2002):
\begin{quote}
\textbf{Original:} ``Television producer by day, CIA assassin by night, \textit{Chuck Barris} was recruited by the CIA at the height of \textit{his} TV career and trained to become a covert operative\ldots''  

\textbf{Anonymized:} ``Television producer by day, CIA assassin by night, \textit{PERSON 1} was recruited by the CIA at the height of \textit{their} TV career and trained to become a covert operative\ldots''.
\end{quote}

% LLM performance is strongly dependent on training data, creating a false impression of its future performance on the same task \cite{chang_speak_2023}, such as future script generation for \textit{non-}recognized movies, for example. 
% We generated LLM scripts based on the anonymized short synopses of the 783 human-written scripts in our dataset. We provided the LLM solely with the movie synopsis to generate its script, which allowed us to study the types of output by the LLM when given creative leeway. Additionally, since the human-written scripts are likely in the training data of the LLM, we anonymized each synopsis to avoid a simple retrieval of the human-written script; we cleaned each synopsis, removing gender-identifying and name-identifying information to minimize the likelihood of the LLM recognizing the movie and/or generating characters based on this recognition. 

\subsubsection{Prompt engineering} 
We conducted pilot experiments to develop an appropriate prompting approach for screenplay generation across GPT, Claude, and Gemini. Previous work that prompted GPT to generate screenplays used a prompt based on a single synopsis, but the resulting scripts were far shorter than human-written scripts \cite{mahomed_auditing_2024}. We determined longer scripts necessary for the representational analysis we focus on in this paper, so length was the focus of our prompt engineering experiments.

We trialed two methods: (1) we first attempted to generate the entire movie screenplay with a single prompt that included the anonymized synopsis, similar to the work of \cite{mahomed_auditing_2024}, and (2) sequential prompting using consecutive paragraphs of the anonymized synopsis, with generated sub-scripts concatenated to form a full screenplay. Neither method generated scripts long enough to be comparable to human-written scripts, although the second was more successful. For both methods, we also experimented with providing a word count target for the output length. However, the LLMs did not respond accurately to this instruction, which aligns with their observed inability to recognize word counts~\cite{Xu_Ma_2025}.

Our final approach successfully generated long screenplays that were of similar or longer length than the human-written counterpart scripts. First, we prompted each model with the anonymized synopsis, specified the target number of scenes, and requested a numbered list of scene headings with brief descriptions. Second, for each scene heading, we issued a follow-up prompt to generate the full scene content. Finally, we concatenated the resulting scenes to form a complete screenplay. The exact prompts used in this two-step approach are shown in Appendices~\ref{fig:screenplay_prompt} and~\ref{fig:scene_prompt}.

\subsubsection{Generated screenplay dataset}
\begin{figure}[!htbp]
    \centering
    \includegraphics[width=0.4\linewidth]{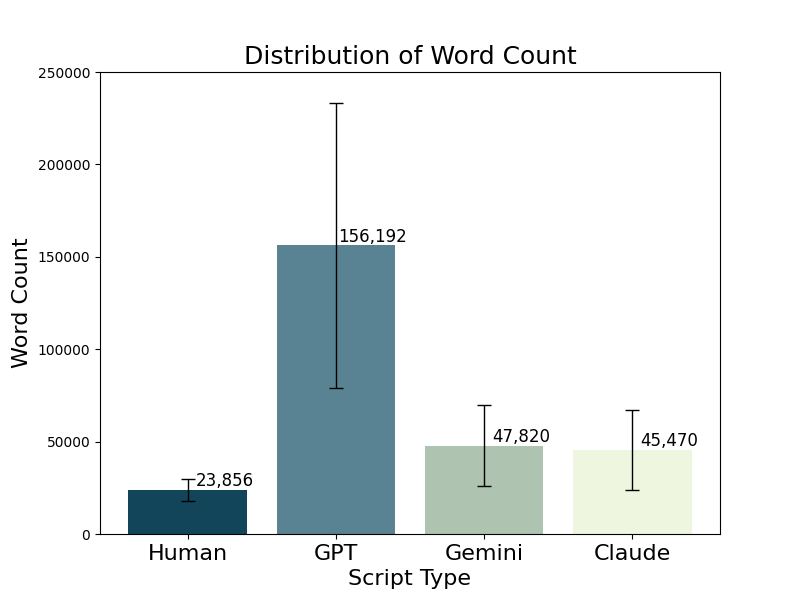}
    \caption{Mean script length across the 4 script types; error bars indicate standard deviation. }
    \label{fig:LLM-word-char-interaction-count}
\end{figure}

We generated 770 GPT scripts ($M=156,192 words, SD=77,030$), 771 Gemini scripts ($M=47,820 words, SD=21,725$), and 752 Claude scripts ($M=45,470 words, SD=21,605$). 

During individual scene generation, models sometimes refused to generate scenes due to content moderation. Claude refused 36 scene generations, while GPT-5 refused twice, each with explicit content-filter errors. Gemini, by contrast, did not issue explicit refusals but instead generated 50 empty scenes. Models also refused or generated empty screenplays with Claude missing 31 screenplays, GPT missing 13 screenplays, and Gemini missing 12 screenplays. We proceed rather than regenerating to remain consistent with a one-time generation per script, and because prior work on refusal behaviors suggests regeneration would likely result in the same outcome.

All three models demonstrated high variation in output length; a given model generated a large range of script lengths. LLM-generated scripts were substantially longer than human-written scripts, with GPT producing scripts $\sim$3.3 times longer than the other LLMs and $\sim$6 times longer than human-written scripts (Figure~\ref{fig:LLM-word-char-interaction-count}). 

Occasionally LLMs reproduced character names despite synopsis anonymization, which may be due to recognizing the source film from training data. Importantly, however, each LLM structures its generated narrative differently, as reflected in varying script lengths and social network structures. In a spot-check reading of 100 screenplays --- 25 films and 4 script types --- we did not find evidence that LLMs simply replicated human-written scripts. Our analysis focuses on these structural differences, rather than whether the original film was recognized. 

\section{Analytical Methods}
We now describe our methods for 
\begin{enumerate*}
    \item constructing character interaction networks from human- and LLM-generated screenplays,
    \item implementing an automated Bechdel test, and
    \item computing network measures to compare structural similarities and differences and to assess gender bias within these networks.
\end{enumerate*}

\subsection{Character Network Construction} 
Characters were extracted and converted to nodes in our network representation.
For both human-written and LLM-generated screenplays, we extracted character names based on formatting patterns: (a) names followed by a colon, (b) names written in all-uppercase, and (c) exclusion of scene directions such as ``FADE IN'' or ``CUT TO.''

Edges connect character nodes to indicate interactions, with edge weights representing the number of times a given pair of characters interacts. Following \citet{agarwal_key_2015}, we define a character interaction as two characters appearing consecutively within the same scene. For example: 

\vspace{-3pt}
\begin{quote}
\texttt{JAMAL:
 Where did you get the gun?}
 
\texttt{SALIM:
 Bought it. Now, I'm going to
 have to throw this beauty in the
 sea.}

\texttt{LATIKA:
 You didn't need to kill him.}
\end{quote}
\vspace{-3pt}

\noindent In this dialogue excerpt from \textit{Slumdog Millionaire} (2008)~\cite{slumdog2008}, the interaction pairs are Jamal/Salim and Salim/Latika. Jamal and Latika are not considered to have interacted, as they did not speak consecutively. This consecutive-speaker approach provides a more accurate representation of character interaction than connecting all characters appearing in the same scene, in keeping with past work~\cite{agarwal_key_2015}.

Using this graph construction method, 768 out of the 783 human-written scripts were able to be converted to character networks (the remaining 15 unconverted scripts lacked scene breaks and other necessary cues). We then generated social networks for the LLM counterparts of the 768 network-convertible human-written scripts. This was possible for all Gemini scripts; five Claude scripts did not contain dialogue, and one GPT script flagged that the prompt asked for too many scenes. This brought our totals to 768 human-written, 754 GPT, 756 Gemini, and 732 Claude character networks. 

\subsubsection{Determining Character Gender}
In the network representation of human-written scripts, we ascribed gender to each character node by matching characters to cast members in TMDB and using the actors' listed genders (no available gender, female, male, or other). Manual review of the ``other'' category revealed inconsistencies: some nodes represented cisgender characters played by non-binary actors, while others were transgender characters played by cisgender actors. Given these inconsistencies and the binary gender framework of the Bechdel test, we reclassified all such nodes, along with `no available gender' nodes, using a name-based approach, classifying character names rather than actor names because actors do not necessarily play characters that correspond to their personal gender identity.  

We used the NomQuamGender~\cite{van_open_2023} Python package to classify unmatched character names in the human-written screenplays and all character names in the LLM-generated ones. NomQuamGender uses a database of names with known gender distributions to calculate the probability of a name belonging to a man or woman based on historical usage patterns, with probability values near 0.5 indicate roughly equal usage across genders. The algorithm applies an uncertainty threshold in classifying names; we tuned NomQuamGender on all character names from TMDB, which determines the smallest uncertainty threshold that can gender-classify >85\% of the character names at a starting threshold of 0.3. The algorithm was maximally able to classify 62\% of the character names at this threshold, so the algorithm proceeded with an uncertainty threshold of 0.3 in our work. The remaining 38\% of the dataset had unclassifiable character names that were role-based (e.g., Therapist, Barista) or gender-neutral (e.g., Ash, Casey).

We acknowledge that inferring gender from names is inherently limited, and risks reinforcing essentialist links between names and gender categories~\cite{gautam_stop_2024}. At best, this is only a proxy: names do not determine gender identity, vary significantly across cultures and time periods, and can be gender-neutral or used across genders. However, for the purposes of Bechdel test analysis---which relies on perceived gender presentation in dialogue---a probabilistic name-based approach provides a consistent method applicable to both human-written and LLM-generated scripts.  

Using this combined approach, we successfully assigned a gender to 70\% of human-written characters. For LLM-generated scripts, gender classification rates were slightly higher: 75\% for GPT, 78\% for Gemini, and 69\% for Claude.  Character roles are determined in Section~\ref{sec:majorchars}.  The majority of unclassified characters are peripheral figures lacking explicit names or gender markers in the script --- such as ``Police Officer'' or ``Teacher'' --- and only 6.90\% of major characters remained unclassified.

\subsubsection{Determining Character Role}
\label{sec:majorchars}
We implement the algorithm developed by~\citet{Park_Oh_Jo_2012} to classify characters into two categories: \textsc{major} and \textsc{non-major} (includes \textsc{minor} and \textsc{peripheral}). This classification is relevant for analyzing film scripts because characters in different roles exert distinct narrative influence. By categorizing characters by role prominence, we can assess whether gender representation varies across these hierarchical levels.

The algorithm uses degree centrality as its primary measure. A character's degree centrality is calculated by dividing its degree (the number of other characters it is connected to) by the maximum possible degree in the network (N-1, where N is the total number of characters). The classification process proceeds in three steps:
\begin{enumerate*}
\item All characters are sorted in descending order by degree centrality. 
\item For each adjacent pair of characters in this sorted list, we calculate the difference in their degree centrality values. 
\item We identify the pair with the largest centrality difference; this difference defines the boundary between \textsc{major} and \textsc{minor} characters. The higher-centrality character in this pair is classified as the last \textsc{major} character, while the lower-centrality character becomes the first \textsc{minor} character. The last minor character has the lowest degree centrality that is still above the mean degree centrality of the network. 
\item All nodes with values below the mean degree centrality are classified as peripheral. However, since we are primarily interested in gendered characters that have influence in the network, we merged the \textsc{minor} and \textsc{peripheral} roles into the singular \textsc{non-major} role.
\end{enumerate*}

\subsection{Automating the Bechdel Test}
% Steph edited 1x
The Bechdel test assesses women's presence and dimensionality in literary or film work through three sequential criteria, each conditional on passing the previous one. Following~\citet{agarwal_key_2015}, we operationalized each criterion:

\textit{Test 1: Two} named \textit{female characters...}
If a movie's character network contained at least two \textit{named} female-classified nodes, the movie passed Test 1. 
% We filtered out common gendered, unnamed character labels like MOTHER, FATHER, NURSE, RECEPTIONIST, and DRIVER (Fig. ~\ref{fig:gendered dictionary}), which were identified through a manual review of detected characters. While this was largely effective, as validated in Section~\ref{methods:validating_bechdel_model}, some nodes still referenced unnamed female characters. 
Otherwise, if the character network had fewer than two named female characters, the movie failed Test 1, and was ineligible to continue the test. 

\textit{Test 2: ...who talk to each other...}
A movie passed Test 2 if its character network contained at least one edge between any two female character nodes. If the network had no edges between female characters, the movie failed Test 2, and was ineligible to continue.

\textit{Test 3: ...about something other than a man.}
To automatically determine whether female-female character conversations were about men, we used a machine learning approach based on SNA features. Following \citet{agarwal_key_2015}, who demonstrated that using the set of SNA features yielded higher F1 scores than alternative feature sets --- including word unigrams, distribution of conversations over topics, linguistic features capturing mentions of ``men'' in dialogue, and features proposed by \citet{garcia_gender_2014} --- we extracted 43 SNA features characterizing the placement of female characters in the character network (see Appendix~\ref{fig:43 SNA features}). We then trained an RBF Support Vector Machine (SVM) model to predict Test 3 scores. The model was trained on 576 randomly selected films (75\% of the 768 human-written films), with hyperparameters optimized via grid search ($C = 0.5$, $\gamma = 0.01$), and subsequently applied to the full dataset of scripts in our testing pipeline. The SVM had a prediction error of $0.265$ and the following accuracy scores: $P = 0.65, R = 0.80, F1 = 0.72$.

\subsubsection{Validation} 
\label{methods:validating_bechdel_model}
We validated our Bechdel test method on human-written scripts using the Bechdel Test Movie List\footnote{https://www.bechdeltest.com/}, a crowdsourced dataset of human evaluations used in the work of \citet{agarwal_key_2015}; our validation also includes films added to the list over the past decade since that publication. Admin approval is required to rate a movie and at the time of data collection, the site had contributions within the same month. In our dataset of 768 movies, 95\% of movies pass the first test, 67\% pass the second, and 54\% pass the third. Note that movies that pass Test 2 are highly likely to also pass Test 3.
 Our method achieved comparable performance to \citet{agarwal_key_2015}, with comparable precision, recall and F1 scores for each test (see Appendix Table{~\ref{tab:bechdel-validation}} for precise numbers).  

Since Test 3 is trained and validated on human-written screenplays, we further assessed robustness under domain shift to LLM-generated screenplays. We hand-annotated 300 LLM screenplays --- 100 each from Gemini, GPT, and Claude --- for pass/fail on the Bechdel test and compared this to our model's assessment. Our model achieves comparable precision, recall, and F1 scores for LLM scripts relative to human-written screenplays, indicating our model generalizes well to LLM-generated screenplays (see Appendix Table{~\ref{tab:llm-validation}} for precise numbers).

 \subsection{Assessing Network Structure}
 \label{methods:comparing_network_structure}
To supplement Bechdel performance, we analyze character centrality, homophily, and triangle relationships as additional measures of gender bias in character interaction and positioning. Below we define and motivate these measures.

\paragraph{Centrality} The centrality of a given character node expresses its influence in the character network; a character with higher involvement with other characters is more central ~\cite{wasserman_social_1994}. There are three main notions of centrality that each offer distinct perspectives:
\begin{enumerate*}
\item \textit{Degree centrality} is based on the idea that a more central character must be connected to more characters in the network, and is defined as a ratio of the number of others to which a given character connects compared to the maximum possible such value~\cite{wasserman_social_1994}.
\item \textit{Closeness centrality} offers an alternate view: a character is more central the faster that node can reach other characters' nodes through the network, calculated by measuring the inverse average distance between a given character $n_i$ and all other characters~\cite{wasserman_social_1994}. 
\item Finally, \textit{betweenness centrality} suggests that a character is more central if their node mediates interactions between non-adjacent characters; that is, that a more central character must lie on more shortest paths between two non-adjacent characters, or the average probability that a given character lies on the shortest path between two other non-adjacent characters~\cite{wasserman_social_1994}.
\end{enumerate*}

\paragraph{Homophily}
The Bechdel test only accounts for female-to-female interactions. To examine cross-gender interactions, we analyze the network's homophily --- the pattern in which members of a social group are more likely to share interaction links with each other than with members of other groups~\cite{mcpherson_birds_2001, Newman_2003}. We examine homophily at the dyadic level, measuring the global tendency for characters to form same-gender versus opposite-gender pairs. This offers a different perspective on the positioning of female characters: high homophily indicates that characters primarily engage in gender-segregated interactions rather than integrating across gender lines within the broader cast. 

\paragraph{Triangles} 
Beyond pairwise connections, we examine how characters cluster into the smallest community structures: triangles (characters connected in a triad). Triangles are fundamental units in social networks, forming the minimal building blocks of community structure~\cite{laniado_gender_2016, Durak_Pinar_Kolda_Seshadhri_2012, granovetter1973strength}.  
While homophily reveals the overall tendency for same-gender connections, triangle analysis reveals the gender composition of these minimal community triads. A network could exhibit moderate homophily but predominantly contain male-majority triangles, indicating that while individual connections may be mixed, actual community structures remain gender-segregated. We examine how the distribution of all-male, all-female, and mixed-gender triangles varies across script types.

\section{Findings}
We now describe the findings from evaluating the four screenplay types: human-written, GPT-5, Gemini 3 Pro, and Claude Sonnet 4.5. 
First, we found that the number of interactions in a network, measured as the sum of edge weights, correlates strongly with script type and script length; LLM scripts were significantly longer and had significantly more interactions. Additionally, our analyses heavily rely on edge-defined relationships; 41 of the 43 SNA features tracked in Test 3 and all following social network analyses are dependent on the edge relationships present in the network. We experimented with additional controls alongside the number of interactions, including total characters, average utterance length, and average number of interactions per character, and found that our trends hold. We therefore control for the number of interactions in all analyses that follow.

% (see Tables~\ref{fig:interactions-by-script} and ~\ref{fig:word-count-interactions}).
Since we aim to identify differences between the four script types on various measures that capture representational bias against women, we primarily use an OLS regression model. This model computes a line of best fit for the four script types and control variable(s) as independent variables relative to a particular measure (dependent variable), to identify correlational relationships. The resulting coefficients and p-values show the correlation (if any) between each script type and a particular bias measure. We run all regressions with human-written scripts as the reference; reported coefficients are relative to a human-written script reference point.
We first report performance on our automated Bechdel test model, then examine SNA measures outlined in ~\ref{methods:comparing_network_structure} to measure biases in their representation of women characters.

\begin{figure}[!htbp]
    \centering
    \begin{minipage}{0.5\textwidth}
        \centering
        \includegraphics[width=1\linewidth]{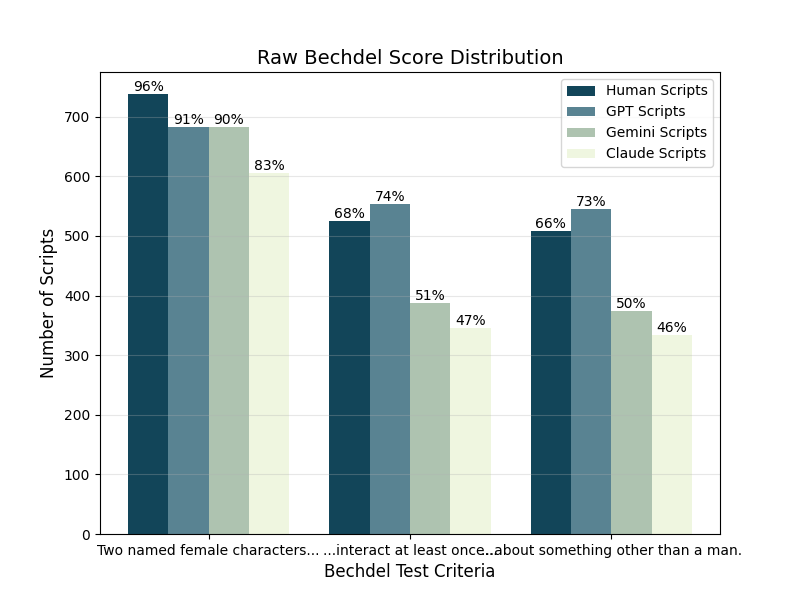}
    \end{minipage}\hfill
    \begin{minipage}{0.5\textwidth}
        \centering
        \includegraphics[width=1\linewidth]{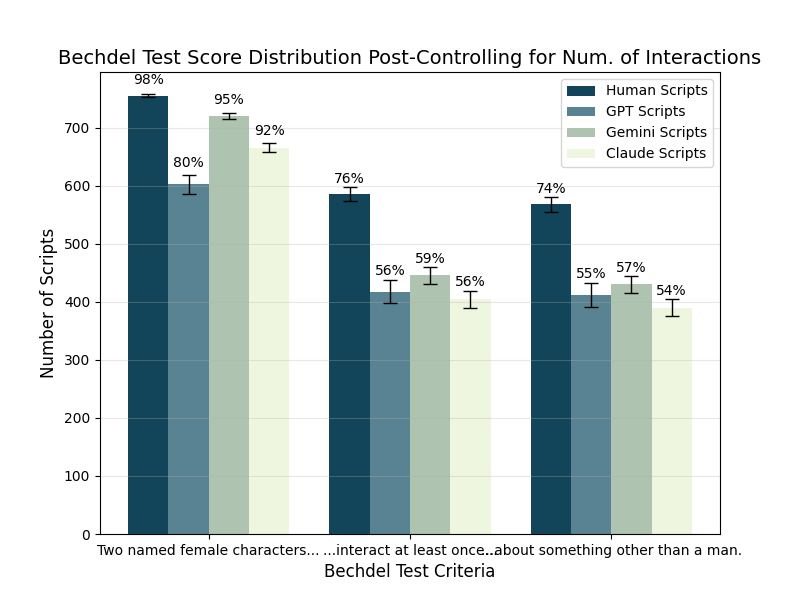}
    \end{minipage}
    \caption{Bechdel test performance before controlling for the number of interactions (left; raw pass rates) and after controlling for interactions (right; predicted pass rates); right error bars indicate predicted rates after controlling for interactions $\pm$ standard error.}
    \label{fig:bechdel-perf}
\end{figure}

\subsection{Human-written screenplays score higher than LLMs on the Bechdel test}
% Steph edited 1x
Based on the three tests described above, each script receives a Bechdel test score from 0 to 3. A script scores 0 if it fails Test 1 (lacks two female characters), 1 if it passes Test 1 but fails Test 2 (female characters do not interact), 2 if it passes Tests 1 and 2 but fails Test 3 (female characters only discuss men), and 3 if it passes all three tests. To pass the Bechdel test overall, a script must achieve a score of 3; scores of 0 to 2 indicate failure.

\paragraph{On a pass/fail basis} Figure~\ref{fig:bechdel-perf} shows 74\% of GPT scripts passed, 66\% of human scripts passed, 50\% of Gemini scripts passed, and 46\% of Claude scripts passed. When we control for the number of interactions \footnote{\texttt{verdict $\sim$ script type $+$ number of interactions}} in a logistic regression model that returns the odds of a specific script type passing the Bechdel test, we find that the number of interactions is correlated with pass/fail outcomes ($\beta = 0.001, p < 0.001$) and all LLM scripts are significantly less likely to pass the Bechdel test than human-written scripts. 

Out of the LLM scripts, Gemini scripts are most likely to pass ($\beta = -0.77, p < 0.001$), followed by GPT scripts ($\beta = -0.88, p < 0.001$), and lastly, Claude scripts ($\beta = -0.91, p < 0.001$). We use the logistic regression model to predict the pass rate at the mean number of interactions, shown in Figure~\ref{fig:bechdel-perf}. This suggests that LLM-generated script performance on the Bechdel test may be inflated due to longer script length and thus more allowance to form interactions.

\paragraph{On an exact score basis} 
After controlling for the number of interactions \footnote{\texttt{score $\sim$ script type $+$ number of interactions}}, all LLM-generated scripts are associated with significantly lower Bechdel scores than human-written scripts. Relative to human-written scripts, Gemini-generated scripts exhibit the smallest decrease in Bechdel score ($\beta = -0.43, p < 0.001$), followed by GPT-generated scripts ($\beta = -0.52, p < 0.001$), while Claude-generated scripts show the largest decrease ($\beta = -0.57, p < 0.001$).

\paragraph{On a differential score basis} 
For each LLM-generated script, we compute the difference in Bechdel score relative to its human-written counterpart as $\text{Score}_{\text{human}} - \text{Score}_{\text{LLM}}$. In an OLS regression model with Claude-generated scripts as the reference category \footnote{\texttt{score difference $\sim$ script type $+$ number of  interactions}}, the score differences for GPT-generated screenplays and Gemini-generated scripts are not statistically significantly different from those for Claude-generated screenplays.  

Across evaluations, human-written screenplays consistently score higher than LLM-generated screenplays on the Bechdel test. In both the binary and continuous score evaluations, Gemini-generated screenplays exhibit the least reduction in performance relative to human-written screenplays, followed by GPT-generated screenplays, with Claude-generated screenplays showing the largest reduction. However, when modeling the Bechdel score gap between human-written and LLM-generated screenplays, differences among LLMs are not statistically significant. We next turn to network-based measures of gender bias. 

\subsection{Some LLM-generated scripts demonstrate better female character representation than human scripts on other network measures}

The Bechdel test is one network evaluation measure of gender bias for film media. To investigate further, we examine several more network metrics, including the centrality imbalance between male and female characters, gender homophily, triadic relationships between male and female characters, and the proportion of edges involving female characters. In doing so, we find a more complex story: some LLM-generated scripts display better women's representation than human ones on other meaningful measures, described below. 

\begin{figure}[t]
    \centering
    \begin{minipage}{0.33\textwidth}
        \includegraphics[width=1\linewidth]{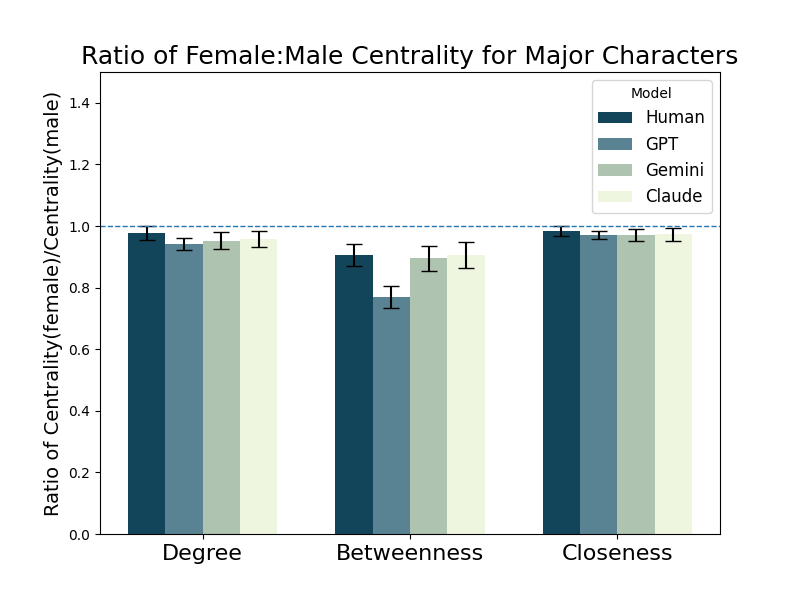}
    \end{minipage}\hfill
    \begin{minipage}{0.33\textwidth}
        \centering
        \includegraphics[width=1\linewidth]{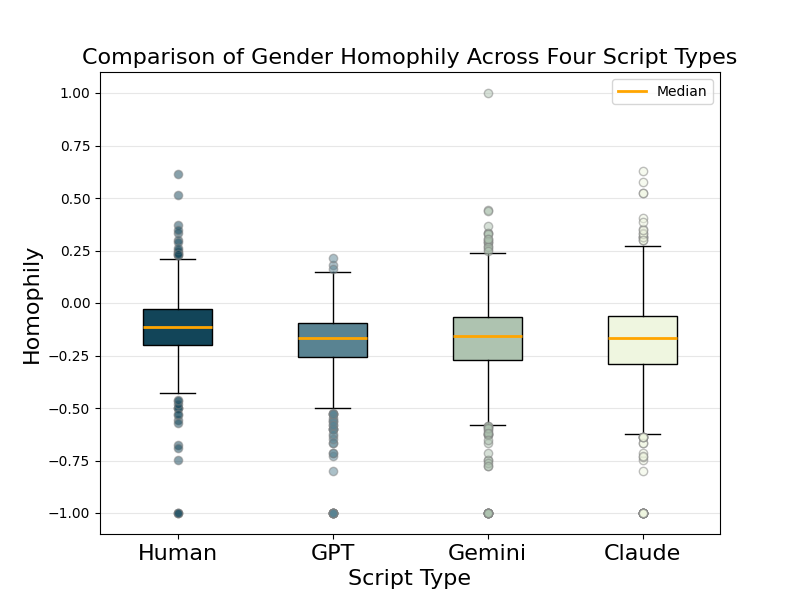}
    \end{minipage}\hfill
    \begin{minipage}{0.33\textwidth}
        \includegraphics[width=1\linewidth]{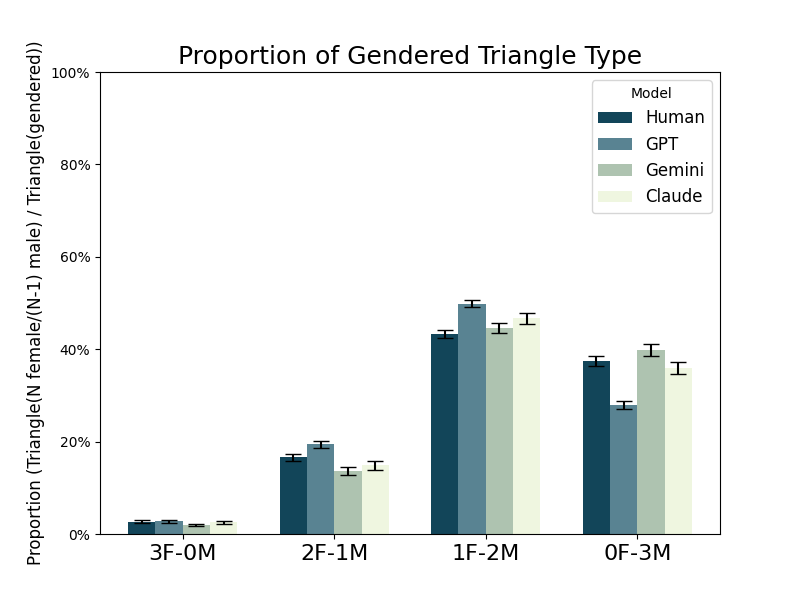}
    \end{minipage}
    \caption{Distribution of various network measures across 4 script types. Left: ratio of female centrality to male centrality for \textsc{major} characters, across all 4 script types and 3 centrality notions. Middle: box plot distribution of gender homophily. Right: proportion of triangles per gender composition type; error bars indicate mean $\pm$ standard error on bar graphs}
    \label{fig:sna-measures}
\end{figure}

\subsubsection{All script types display gender bias in \textsc{major} character centrality}
%\subsubsection{Claude screenplays exhibit the highest relative female centrality, whereas GPT screenplays exhibit the lowest}

The Bechdel test treats all character interactions equally and does not account for differences in character role. 
% This approach overlooks the systematically different influence that each role type has on a film’s narrative structure \cite{park_character-net_2009}.
Accordingly, we examined gender differences in centrality for \textsc{major} characters (see Section~\ref{sec:majorchars} for definition and construction). For each of the three notions of centrality previously defined, we conducted pairwise t-tests to assess differences in centrality between male and female characters.

% \subsubsection{\textsc{Major} character centrality}

All scripts display gender bias towards female characters, as seen in the ratio of female to male centrality falling under 1 in Figure \ref{fig:sna-measures}. However, a statistical difference in centrality between male and female characters is only seen in GPT and human-written scripts, with GPT scripts demonstrating the most uneven representation of female \textsc{major} characters. 
GPT-generated screenplays exhibited a statistically significant difference between the mean degree centrality of male ($mean = 0.61$) and female ($mean = 0.58$) \textsc{major} characters, with male \textsc{major} characters more central ($t = 2.68, p = 0.007$). In contrast, human-written, Claude, and Gemini scripts show no significant difference in mean degree centrality between genders.
% (Fig. \ref{fig:major-centrality}).
%
Only GPT ($t = 5.72, p < 0.001$) and human-written ($t = 2.42, p = 0.016$) scripts demonstrated a higher betweenness centrality for male than female \textsc{major} characters. GPT scripts exhibited the larger betweenness centrality gap between male and female \textsc{major} characters, as shown by the smaller ratio of female-to-male betweenness centrality ($ratio_{GPT} = 0.78, ratio_{Human} = 0.91$).
% (See fig. \ref{fig:major-centrality}). 
%
We did not find statistical differences between genders for closeness centrality across script types.

\subsubsection{GPT screenplays exhibit the lowest homophily of female characters}
All scripts displayed heterophily (the inverse of homophily) with mean scores from $-0.11$ to $-0.19$, as seen in Figure{~\ref{fig:sna-measures}}, reflecting more inter-gender interactions than expected. (Unlike movie scripts, which establish connection strictly through dialogue, real-world relationships are defined more robustly and tend to be more homophilous.) In an OLS regression model \footnote{\texttt{homophily $\sim$ script type $+$ number of interactions}}, GPT scripts displayed the highest heterophily ($\beta = -0.13, p < 0.001$), followed by Claude scripts ($\beta = -0.08, p < 0.001$), Gemini scripts ($\beta = -0.07, p < 0.001$), and lastly human-written scripts. 
This suggests that while human-written and Gemini scripts score higher than GPT and Claude scripts on the Bechdel test, this may be a result of better interaction between female characters than cross-gender interactions with male characters. 

\subsubsection{GPT screenplays have more female-inclusive triadic relationships than others}
The second Bechdel test is a measure of female-female dyads in a network (two women that have a conversation). We examine a stronger bar: triads, network segments involving three characters.
We examine four types of triadic relationships, classified by gender composition: three female and zero male (3F–0M), two female (2F–1M), two male (1F–2M), and three male characters (0F–3M) (Figure \ref{fig:sna-measures}). Across all script types, the most common triangle is 1F–2M, followed by 0F–3M, 2F–1M, and finally 3F–0M. These patterns are consistent with previous observations in human-written film screenplays~\cite{kagan_using_2020}. 
Aggregating triangles with at least one female character as \textit{female-inclusive}, and triangles with only male characters as \textit{female-exclusive}\footnote{\texttt{proportion of female-inclusive triangles $\sim$ script type + number of interactions}}, GPT scripts exhibited a significantly higher proportion of female-inclusive triangles ($\beta = 0.09, p < 0.001$) compared to the other three types, which performed similarly. 
% By this measure, GPT scripts perform the best overall, with the highest proportion of female-inclusive triadic relationships. Gemini scripts performed the worst by exhibiting the lowest proportion of two female and two male relationships and highest proportion of three male.

% , which contradicts its relatively higher performance on the Bechdel test. 

% This further supports the above conclusion that GPT scripts contain the most heterophily. Claude scripts and human-written scripts performed similarly, indicating...

\begin{figure}[!t]
    \centering
    \includegraphics[width=0.4\linewidth]{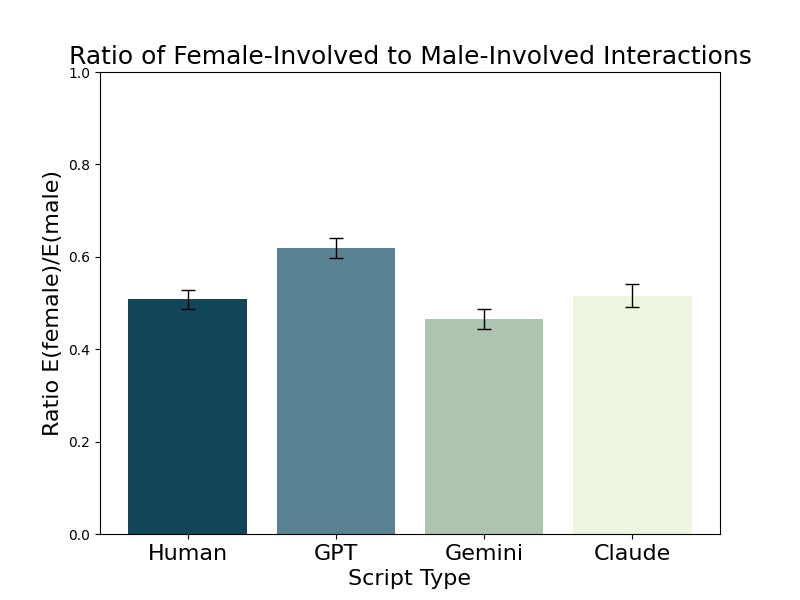}
\caption{Ratio of female-involved interactions (edges) to male-involved interactions; error bars indicate mean $\pm$ standard error.}
\label{fig:major-centrality-dist}
\end{figure}

\subsubsection{GPT screenplays have the highest proportion of interactions involving female characters}
We analyze representational bias in films by examining both the proportion of female characters and the proportion of interactions involving them. While the Bechdel test incorporates the proportion of female characters, it does not distinguish characters by role type or account for the ratio of interactions in which female characters participate.
We examine the proportion of female characters among \textsc{major} characters as defined in prior work; these characters play a central role in the narrative structure of films \cite{park_character-net_2009}. Using an OLS regression model controlling for number of character interactions \footnote{\texttt{proportion of \textsc{MAJOR} female characters $\sim$ script type + number of  interactions}}, we find no statistically significant differences across script types: all contain roughly similar proportions of female \textsc{major} characters, ranging from 21–23\% (n.s.).
Examining the ratio of interactions involving female versus male characters with another regression \footnote{\texttt{ratio of female-involved interactions to male-involved interactions $\sim$ script type + number of  interactions}}, GPT scripts show the highest ratio ($\beta = 0.11, p < 0.001$), while human-written, Gemini ($\beta = -0.04, p = n.s.$), and Claude scripts ($\beta = 0.01, p = n.s.$) do not differ significantly, as shown in Figure \ref{fig:major-centrality-dist}.

\section{Discussion}
We find that human screenplays score higher than LLM-generated ones on the Bechdel test, a popular measure of gender bias in film. Our findings are more mixed on other aspects of their character networks; GPT-generated scripts display better representation of women than human ones on some measures, including all-female network triads and heterophily. We next discuss implications for future studies, and the broader risks of AI replacing human creative labor. 

\subsection{Expanding LLM Evaluation Strategies}
% DM: apologies sorelle this is still really rough
The introduction of LLMs and their surge in popularity has led to a parallel explosion in studies auditing and evaluating LLM outputs for bias \cite{abid_persistent_2021, kirk_bias_2021, liu_racial_2024, lucy_gender_2021, mahomed_auditing_2024, metaxa_auditing_2021, sheng_woman_2019, wan_kelly_2023}. Auditing AI systems is critical for raising awareness among users about potential harms, and seeking accountability from developers. In that spirit, in this work we sought to analyze representational fairness \cite{barocas_problem_2017, blodgett_language_2020, crawford_trouble_2017} in LLM depictions of women. 
Conducting such representational bias evaluations entails several challenges compared to studying questions of allocative fairness, first among them the unstructured nature of their outputs. The long-form text produced by LLMs is hard to quantitatively analyze, leading many bias auditors to instead force LLMs to give binary or numerical answers \cite{wilson_gender_2024, Iso_Pezeshkpour_Bhutani_Hruschka_2025}, or to extract and analyze LLMs' internal vector representations of different concepts \cite{Cyberey_Ji_Evans_2025}. While such audits are useful, they necessarily remove some of the nuance of generative textual responses.

% Most ML fairness work has conflated media representations with internal LLM vector representations; LLM vector representations are the main way that people think about representational fairness. But we want to go beyond that, while still allowing structured audits and quantification. Social network analysis is one way to do this. 

In this work, we take a complementary approach drawing from social network analysis (SNA), converting LLM outputs into network graphs that can be quantitatively analyzed using SNA measures. This has the added benefit of allowing us to compare LLM outputs with human texts, as such analyses are commonly done on human-written content in the domain of film \cite{agarwal_parsing_2014, agarwal_key_2015, kagan_using_2020, kumar_gender_2022, lv_storyrolenet_2018, park_character-net_2009}. 
Our findings demonstrate the value of a network analysis approach to this topic, and support its broader use in LLM evaluations, including across other demographic groups and/or forms of media beyond screenplays. More generally, we encourage researchers in this domain to identify other ways to impose structure and leverage proven techniques to develop new forms of LLM evaluation. 

\subsection{Beyond Bechdel and Bias}

A full and nuanced sense of self-identification and representation in media --- the sense of seeing yourself in a character --- cannot be captured by any quantitative measure. The same is true for positive or equitable representation of women (or any other group). In this work, we examine a variety of measures that capture different dimensions of visual representation --- centrality within the narrative, tendency to form cross-gender relationships, proportion of characters of the same gender, etc. --- to capture some aspect of these social concepts. But even if LLM scripts could match or outperform humans on our metrics, this would neither give people a sense of genuine visual representation, nor ensure fairness or equity in that representation.

Beyond representation, another crucial issue is at stake: the treatment of the human artists whose work is increasingly ingested, often without permission, to power systems that attempt to replace them. 
Alongside research on LLM biases in creative tasks, more attention is needed to the experiences of creatives on the front lines of this shift. Such questions complement, and reinforce, questions of representation in media; screenwriters should have a voice in determining both whether and how LLM screenwriting tools are used, and in assessing the quality of any resulting portrayals of the human experience. Participatory methods, such as those described by \citet{tseng_ownership_2025} in their co-design of a journalist-controlled LLM are promising paths forward. By involving stakeholders whose livelihoods are most affected, like journalists and screenwriters, as co-designers, these approaches help preserve their decision-making power.

% - dont lose sight of the forest for the trees, even if we are just studying the trees
% - also more work needed studying the forest

\subsection{Limitations}
% DM edited, happy with it
As mentioned in previous sections, this work entails several limitations worth discussion.
% \subsubsection{Name-to-gender inferences}
The first is that our work relies on name-to-gender inference to enable our large-scale analysis of gender bias in screenplays, which are dialogue-heavy and lack sufficient pronoun references. Automated name annotation itself introduces error, names are imperfect proxies for identity, and this method limits our work to a binary framing. Further work is sorely needed to address these issues, especially as characters beyond the binary are increasingly being represented in film. 

% \subsubsection{Dataset characteristics}
We also collected scripts and Bechdel scores from crowdsourced datasets. Though these sites are moderated, their data can be highly variable. For example, 45\% of the human-written scripts were not tagged with maturity ratings and each film is tagged with multiple genres. This limitation extends to LLM-generated scripts, which similarly lack genre/rating classifications, so we were unable to segment our analyses along these axes. Bechdel scores were also imperfect; when manually reviewing score disagreement between our model and the human-annotated baseline (from The Bechdel Movie List), we found several instances where the latter had inconsistently applied the test and/or users disagreed on the score. Additionally, we acknowledge that LLMs are non-deterministic. We believe our full dataset of over 700 screenplays is robust, but further work could be done on model consistency.

% \subsubsection{LLM-specific limitations}  
Finally, since we collected scripts from publicly available sources, there is a high likelihood that the scripts appear in the LLMs' training data despite our attempts at anonymization. 
% \hl{Our synopsis prompting process resembles a name cloze task --- an evaluation used to measure exact memorization in LLMs by prompting them to fill in blanks in anonymized text excerpts \cite{Chang_Cramer_Soni_Bamman_2023}. While we do not employ a name cloze task, the resemblance to our synopsis anonymization process and occurence of similar characters across scripts implies that LLMs largely recognize movies in our work}. 
However, given that the resulting scripts were substantially longer than the human-written versions, had structurally different social networks, and did not replicate human scripts during a spot check reading of 100 screenplays, we do not believe the generated scripts directly reproduced such training data. Future work could further investigate the extent to which the generated scripts represented new plot ideas and different character development. 

\section{Conclusion}

LLMs have well-documented biases against various marginalized groups, including women, but their long-form outputs make these challenging to study quantitatively. Existing analyses have largely been restricted to studying internal vector representations or constraining LLM outputs to single-word or otherwise structured answers. In this work, we complement these approaches in a timely creative domain, 
applying social network analysis to analyze LLM-generated screenplays. We curate a dataset of 768 human-written scripts and create matching LLM-generated scripts by GPT, Gemini, and Claude based on the same film synopses.
We create character networks for each script and analyze the network structures in two parts: (1) using the Bechdel test, a popular measure of gender bias in film, and (2) analyzing the representation of female characters using network measures informed by the literature, including centrality and homophily.
We find that human scripts outperform generated ones on the Bechdel test, but that LLM-generated scripts display less bias against women in some of the subsequent tests. We find that across tests and script type, representational bias against women remains.

Based on these findings, we discuss implications for creative fields like filmmaking, and for researchers studying fairness and accountability.  
% DM: I dont' really like the below anymore, bleh, but won't have time to fix
These findings highlight the value of LLM audits for evaluating potential biases, and the importance of developing methodological tools -- such as the social network analyses performed here -- for conducting them.  
Representational bias is particularly concerning as AI-generated text, images, and video increasingly shape our media environment. While human-produced media has long reflected structural biases, frameworks such as the Bechdel test have helped make these patterns visible and have contributed to gradual improvements in representation. Just as human writers’ rooms can bring in consultants or screenwriters from diverse backgrounds, affected stakeholders should likewise be involved as co-designers in determining whether and how LLMs are used in creative production.

% Representational biases are especially concerning as AI-generated text, image, and video increasingly shapes our media environment. While human-generated media is also known to be biased, media critiques like the Bechdel test have shed light on this bias, and representation has improved gradually over time. Unlike human writers' rooms, where consultants, representatives, or screenwriters from marginalized backgrounds can be brought in to offer diverse perspectives, LLMs risk remaining stuck in a media past. 

%As the movie industry moves towards incorporating generative AI in all stages of the production workflow including ideation, screenwriting, and editing, LLMs and other forms of AI threaten to replace human contributors. Bias is just one of many potential harms that this threat poses. 

\section*{Statement on Generative AI Usage}
The authors disclose that generative AI, beyond being used as an object of study, aided in coding figures in Python, translating pseudocode into executable Python code, and rephrasing already-written text for clarity.

% \begin{acks}
% ACKNOWLEDGEMENTS
% \end{acks}

\bibliographystyle{ACM-Reference-Format}
\bibliography{bibliography}

@misc{toonkel_openai_2025,
	title = {Exclusive {\textbar} {OpenAI} {Backs} {AI}-{Made} {Animated} {Feature} {Film}},
    author = {Toonkel, Jessica},
    month = sep,
    year = {2025},
	url = {https://www.wsj.com/tech/ai/openai-backs-ai-made-animated-feature-film-389f70b0},
    journal = {Wall Street Journal},
	urldate = {2026-01-01},
}

@misc{glaad2023,
    author = {Joel Kim Booster},
    title = {GLAAD Media Awards 2023: Fire Island's Joel Kim Stands Strong With WGA In Acceptance Speech: ``Labor Issues Are Queer Issues''},
    howpublished = {\url{https://glaad.org/glaad-media-awards-2023-fire-islands-joel-kim-stands-strong-wga-acceptance-speech-labor-issues/}},
    month = {May 14,},
    year = {2023}
}

@article{granovetter1973strength,
  title={The strength of weak ties},
  author={Granovetter, Mark S},
  journal={American journal of sociology},
  volume={78},
  number={6},
  pages={1360--1380},
  year={1973},
  publisher={University of Chicago Press}
}

@article{lee_lionsgate_2024,
	chapter = {Film},
	title = {Lionsgate partners with {AI} firm to train generative model on film and {TV} library},
	issn = {0261-3077},
	url = {https://www.theguardian.com/film/2024/sep/18/lionsgate-ai},
	language = {en-GB},
	urldate = {2026-01-01},
	journal = {The Guardian},
	author = {Lee, Benjamin},
	month = sep,
	year = {2024},
}

@misc{kinder_hollywood_nodate,
	title = {Hollywood writers went on strike to protect their livelihoods from generative {AI}. {Their} remarkable victory matters for all workers.},
	url = {https://www.brookings.edu/articles/hollywood-writers-went-on-strike-to-protect-their-livelihoods-from-generative-ai-their-remarkable-victory-matters-for-all-workers/},
	language = {en-US},
	urldate = {2026-01-01},
	journal = {Brookings},
	author = {Kinder, Molly},
    month = apr,
    year = {2024},
}

@article{chang_llms_2025,
	title = {{LLMs} {Generate} {Structurally} {Realistic} {Social} {Networks} but {Overestimate} {Political} {Homophily}},
	volume = {19},
	copyright = {Copyright (c) 2025 Association for the Advancement of Artificial Intelligence},
	issn = {2334-0770},
	url = {https://ojs.aaai.org/index.php/ICWSM/article/view/35820},
	doi = {10.1609/icwsm.v19i1.35820},
	language = {en},
	urldate = {2026-01-01},
	journal = {Proceedings of the International AAAI Conference on Web and Social Media},
	author = {Chang, Serina and Chaszczewicz, Alicja and Wang, Emma and Josifovska, Maya and Pierson, Emma and Leskovec, Jure},
	month = jun,
	year = {2025},
	pages = {341--371},
}

@article{metaxa_auditing_2021,
	title = {Auditing {Algorithms}: {Understanding} {Algorithmic} {Systems} from the {Outside} {In}},
	volume = {14},
	issn = {1551-3955, 1551-3963},
	shorttitle = {Auditing {Algorithms}},
	url = {http://www.nowpublishers.com/article/Details/HCI-083},
	doi = {10.1561/1100000083},
	language = {en},
	number = {4},
	urldate = {2026-01-01},
	journal = {Foundations and Trends® in Human–Computer Interaction},
	author = {Metaxa, Danaë and Park, Joon Sung and Robertson, Ronald E. and Karahalios, Karrie and Wilson, Christo and Hancock, Jeff and Sandvig, Christian},
	year = {2021},
	pages = {272--344},
}

@article{bamman_measuring_2024,
	title = {Measuring diversity in {Hollywood} through the large-scale computational analysis of film},
	volume = {121},
	url = {https://www.pnas.org/doi/10.1073/pnas.2409770121},
	doi = {10.1073/pnas.2409770121},
	number = {46},
	urldate = {2025-12-31},
	journal = {Proceedings of the National Academy of Sciences},
	author = {Bamman, David and Samberg, Rachael and So, Richard Jean and Zhou, Naitian},
	month = nov,
	year = {2024},
	note = {Publisher: Proceedings of the National Academy of Sciences},
	pages = {e2409770121},
}

@inproceedings{agarwal_key_2015,
	address = {Denver, Colorado},
	title = {Key {Female} {Characters} in {Film} {Have} {More} to {Talk} {About} {Besides} {Men}: {Automating} the {Bechdel} {Test}},
	shorttitle = {Key {Female} {Characters} in {Film} {Have} {More} to {Talk} {About} {Besides} {Men}},
	url = {https://aclanthology.org/N15-1084/},
	doi = {10.3115/v1/N15-1084},
	urldate = {2025-12-31},
	booktitle = {Proceedings of the 2015 {Conference} of the {North} {American} {Chapter} of the {Association} for {Computational} {Linguistics}: {Human} {Language} {Technologies}},
	publisher = {Association for Computational Linguistics},
	author = {Agarwal, Apoorv and Zheng, Jiehan and Kamath, Shruti and Balasubramanian, Sriramkumar and Ann Dey, Shirin},
	editor = {Mihalcea, Rada and Chai, Joyce and Sarkar, Anoop},
	month = may,
	year = {2015},
	pages = {830--840},
}

@article{garcia_gender_2014,
	title = {Gender {Asymmetries} in {Reality} and {Fiction}: {The} {Bechdel} {Test} of {Social} {Media}},
	volume = {8},
	copyright = {Copyright (c) 2021 Proceedings of the International AAAI Conference on Web and Social Media},
	issn = {2334-0770},
	shorttitle = {Gender {Asymmetries} in {Reality} and {Fiction}},
	url = {https://ojs.aaai.org/index.php/ICWSM/article/view/14522},
	doi = {10.1609/icwsm.v8i1.14522},
	language = {en},
	number = {1},
	urldate = {2026-01-01},
	journal = {Proceedings of the International AAAI Conference on Web and Social Media},
	author = {Garcia, David and Weber, Ingmar and Garimella, Venkata},
	month = may,
	year = {2014},
	pages = {131--140},
}

@inproceedings{lucy_gender_2021,
	address = {Virtual},
	title = {Gender and {Representation} {Bias} in {GPT}-3 {Generated} {Stories}},
	url = {https://aclanthology.org/2021.nuse-1.5/},
	doi = {10.18653/v1/2021.nuse-1.5},
	urldate = {2026-01-01},
	booktitle = {Proceedings of the {Third} {Workshop} on {Narrative} {Understanding}},
	publisher = {Association for Computational Linguistics},
	author = {Lucy, Li and Bamman, David},
	editor = {Akoury, Nader and Brahman, Faeze and Chaturvedi, Snigdha and Clark, Elizabeth and Iyyer, Mohit and Martin, Lara J.},
	month = jun,
	year = {2021},
	pages = {48--55},
}

@inproceedings{mahomed_auditing_2024,
	address = {New York, NY, USA},
	series = {{FAccT} '24},
	title = {Auditing {GPT}'s {Content} {Moderation} {Guardrails}: {Can} {ChatGPT} {Write} {Your} {Favorite} {TV} {Show}?},
	isbn = {979-8-4007-0450-5},
	shorttitle = {Auditing {GPT}'s {Content} {Moderation} {Guardrails}},
	url = {https://doi.org/10.1145/3630106.3658932},
	doi = {10.1145/3630106.3658932},
	urldate = {2025-12-31},
	booktitle = {Proceedings of the 2024 {ACM} {Conference} on {Fairness}, {Accountability}, and {Transparency}},
	publisher = {Association for Computing Machinery},
	author = {Mahomed, Yaaseen and Crawford, Charlie M. and Gautam, Sanjana and Friedler, Sorelle A. and Metaxa, Danaë},
	month = jun,
	year = {2024},
	pages = {660--686},
}

@book{bechdel_dykes_1985,
	title = {Dykes to {Watch} {Out} {For}},
	url = {https://dykestowatchoutfor.com/},
    author = {Bechdel, Alison},
    year = {1985},
    publisher = {Firebrand Books},
	urldate = {2026-01-01},
}

@book{kellner_media_1995,
	series = {Media culture},
	title = {Media {Culture}: {Cultural} {Studies}, {Identity} and {Politics} {Between} the {Modern} and the {Postmodern}},
	isbn = {978-0-415-10570-5},
	url = {https://books.google.com/books?id=GjbdsiZ0q10C},
	publisher = {Routledge},
	author = {Kellner, D.},
	year = {1995},
	lccn = {94007262},
}

@article{smith_inequality_2023,
	title = {Inequality in 1,600 popular films: Examining Portrayals of Gender, Race/Ethnicity, LGBTQ+ \& Disability from 2007 to 2022},
	language = {en},
    month = aug,
    year = {2023},
	author = {Smith, Dr Stacy L and Pieper, Dr Katherine and Wheeler, Sam},
    url = {https://assets.uscannenberg.org/docs/aii-inequality-in-1600-popular-films-20230811.pdf},
}

@inproceedings{abid_persistent_2021,
	address = {New York, NY, USA},
	series = {{AIES} '21},
	title = {Persistent {Anti}-{Muslim} {Bias} in {Large} {Language} {Models}},
	isbn = {978-1-4503-8473-5},
	url = {https://doi.org/10.1145/3461702.3462624},
	doi = {10.1145/3461702.3462624},
	urldate = {2026-01-02},
	booktitle = {Proceedings of the 2021 {AAAI}/{ACM} {Conference} on {AI}, {Ethics}, and {Society}},
	publisher = {Association for Computing Machinery},
	author = {Abid, Abubakar and Farooqi, Maheen and Zou, James},
	month = jul,
	year = {2021},
	pages = {298--306},
}

@inproceedings{sheng_woman_2019,
	address = {Hong Kong, China},
	title = {The {Woman} {Worked} as a {Babysitter}: {On} {Biases} in {Language} {Generation}},
	shorttitle = {The {Woman} {Worked} as a {Babysitter}},
	url = {https://aclanthology.org/D19-1339/},
	doi = {10.18653/v1/D19-1339},
	urldate = {2026-01-03},
	booktitle = {Proceedings of the 2019 {Conference} on {Empirical} {Methods} in {Natural} {Language} {Processing} and the 9th {International} {Joint} {Conference} on {Natural} {Language} {Processing} ({EMNLP}-{IJCNLP})},
	publisher = {Association for Computational Linguistics},
	author = {Sheng, Emily and Chang, Kai-Wei and Natarajan, Premkumar and Peng, Nanyun},
	month = nov,
	year = {2019},
	pages = {3407--3412},
}

@article{hofmann_ai_2024,
	title = {{AI} generates covertly racist decisions about people based on their dialect},
	volume = {633},
	copyright = {2024 The Author(s)},
	issn = {1476-4687},
	url = {https://www.nature.com/articles/s41586-024-07856-5},
	doi = {10.1038/s41586-024-07856-5},
	language = {en},
	number = {8028},
	urldate = {2026-01-01},
	journal = {Nature},
	author = {Hofmann, Valentin and Kalluri, Pratyusha Ria and Jurafsky, Dan and King, Sharese},
	month = sep,
	year = {2024},
	note = {Publisher: Nature Publishing Group},
	pages = {147--154},
}

@inproceedings{wan_kelly_2023,
	address = {Singapore},
	title = {“{Kelly} is a {Warm} {Person}, {Joseph} is a {Role} {Model}”: {Gender} {Biases} in {LLM}-{Generated} {Reference} {Letters}},
	shorttitle = {“{Kelly} is a {Warm} {Person}, {Joseph} is a {Role} {Model}”},
	url = {https://aclanthology.org/2023.findings-emnlp.243/},
	doi = {10.18653/v1/2023.findings-emnlp.243},
	urldate = {2026-01-03},
	booktitle = {Findings of the {Association} for {Computational} {Linguistics}: {EMNLP} 2023},
	publisher = {Association for Computational Linguistics},
	author = {Wan, Yixin and Pu, George and Sun, Jiao and Garimella, Aparna and Chang, Kai-Wei and Peng, Nanyun},
	editor = {Bouamor, Houda and Pino, Juan and Bali, Kalika},
	month = dec,
	year = {2023},
	pages = {3730--3748},
}

@article{wang_large_2025,
	title = {Large language models that replace human participants can harmfully misportray and flatten identity groups},
	volume = {7},
	copyright = {2025 The Author(s), under exclusive licence to Springer Nature Limited},
	issn = {2522-5839},
	url = {https://www.nature.com/articles/s42256-025-00986-z},
	doi = {10.1038/s42256-025-00986-z},
	language = {en},
	number = {3},
	urldate = {2026-01-03},
	journal = {Nature Machine Intelligence},
	author = {Wang, Angelina and Morgenstern, Jamie and Dickerson, John P.},
	month = mar,
	year = {2025},
	note = {Publisher: Nature Publishing Group},
	pages = {400--411},
}

@inproceedings{kirk_bias_2021,
	address = {Red Hook, NY, USA},
	series = {{NIPS} '21},
	title = {Bias out-of-the-box: an empirical analysis of intersectional occupational biases in popular generative language models},
	isbn = {978-1-7138-4539-3},
	shorttitle = {Bias out-of-the-box},
	urldate = {2026-01-03},
	booktitle = {Proceedings of the 35th {International} {Conference} on {Neural} {Information} {Processing} {Systems}},
	publisher = {Curran Associates Inc.},
	author = {Kirk, Hannah Rose and Jun, Yennie and Iqbal, Haider and Benussi, Elias and Volpin, Filippo and Dreyer, Frederic A. and Shtedritski, Aleksandar and Asano, Yuki M.},
	month = dec,
	year = {2021},
	pages = {2611--2624},
}

@inproceedings{blodgett_language_2020,
	address = {Online},
	title = {Language ({Technology}) is {Power}: {A} {Critical} {Survey} of “{Bias}” in {NLP}},
	shorttitle = {Language ({Technology}) is {Power}},
	url = {https://aclanthology.org/2020.acl-main.485/},
	doi = {10.18653/v1/2020.acl-main.485},
	urldate = {2026-01-04},
	booktitle = {Proceedings of the 58th {Annual} {Meeting} of the {Association} for {Computational} {Linguistics}},
	publisher = {Association for Computational Linguistics},
	author = {Blodgett, Su Lin and Barocas, Solon and Daumé III, Hal and Wallach, Hanna},
	editor = {Jurafsky, Dan and Chai, Joyce and Schluter, Natalie and Tetreault, Joel},
	month = jul,
	year = {2020},
	pages = {5454--5476},
}

@inproceedings{barocas_problem_2017,
  title={The problem with bias: From allocative to representational harms in machine learning},
  author={Barocas, Solon and Crawford, Kate and Shapiro, Aaron and Wallach, Hanna},
  booktitle={SIGCIS conference paper},
  year={2017}
}

@inproceedings{crawford_trouble_2017,
  title={The {{Trouble}} with {{Bias}}},
  author={Crawford, Kate},
  booktitle={Keynote at NeurIPS},
  year={2017}
}

@article{mccabe_gender_2011,
 ISSN = {08912432},
 URL = {http://www.jstor.org/stable/23044136},
 author = {McCabe, Janice and Fairchild, Emily  and Grauerholz, Liz and A. Pescosolido, Bernice  and Tope, Daniel},
 journal = {Gender and Society},
 number = {2},
 pages = {197--226},
 publisher = {Sage Publications, Inc.},
 title = {Gender in the Twentieth-Century Children's Books: Patterns of Disparity in Titles and Central Characters},
 urldate = {2026-01-03},
 volume = {25},
 year = {2011}
}

@article{tukachinsky_documenting_2015,
	title = {Documenting {Portrayals} of {Race}/{Ethnicity} on {Primetime} {Television} over a 20-{Year} {Span} and {Their} {Association} with {National}-{Level} {Racial}/{Ethnic} {Attitudes}},
	url = {https://digitalcommons.chapman.edu/comm_articles/17},
	doi = {10.1111/josi.12094},
	journal = {Communication Faculty Articles and Research},
	author = {Tukachinsky, Riva and Mastro, Dana and Yarchi, Moran},
	month = jan,
	year = {2015},
}

@inproceedings{van_open_2023,
  title={An Open-Source Cultural Consensus Approach to Name-Based Gender Classification},
  author={Van Buskirk, Ian and Clauset, Aaron and Larremore, Daniel B},
  booktitle={Proceedings of the International AAAI Conference on Web and Social Media},
  volume={17},
  pages={866--877},
  note = {\url{https://github.com/ianvanbuskirk/nbgc}},
  year={2023}
}

@inproceedings{park_character-net_2009,
	title = {Character-{Net}: {Character} {Network} {Analysis} from {Video}},
	volume = {1},
	shorttitle = {Character-{Net}},
	url = {https://ieeexplore.ieee.org/document/5286057},
	doi = {10.1109/WI-IAT.2009.54},
	urldate = {2026-01-06},
	booktitle = {2009 {IEEE}/{WIC}/{ACM} {International} {Joint} {Conference} on {Web} {Intelligence} and {Intelligent} {Agent} {Technology}},
	author = {Park, Seung-Bo and Kim, Yoo-Won and Uddin, Mohammed Nazim and Jo, Geun-Sik},
	month = sep,
	year = {2009},
	pages = {305--308},
}

@misc{bailyn_top_2025,
	title = {Top {Generative} {AI} {Chatbots} by {Market} {Share} – {December} 2025},
	url = {https://firstpagesage.com/reports/top-generative-ai-chatbots/},
	language = {en-US},
	urldate = {2026-01-03},
	journal = {First Page Sage},
	author = {Bailyn, Evan},
	month = dec,
	year = {2025},
	note = {Section: SEO Blog},
}

@article{papachristou_network_2025,
	title = {Network formation and dynamics among multi-{LLMs}},
	volume = {4},
	issn = {2752-6542},
	url = {https://doi.org/10.1093/pnasnexus/pgaf317},
	doi = {10.1093/pnasnexus/pgaf317},
	number = {12},
	urldate = {2026-01-07},
	journal = {PNAS Nexus},
	author = {Papachristou, Marios and Yuan, Yuan},
	month = dec,
	year = {2025},
	pages = {pgaf317},
}

@article{saxena_fairsna_2024,
	title = {{FairSNA}: {Algorithmic} {Fairness} in {Social} {Network} {Analysis}},
	copyright = {Permission to make digital or hard copies of all or part of this work for personal or classroom use is granted without fee provided that copies are not made or distributed for profit or commercial advantage and that copies bear this notice and the full citation on the first page. Copyrights for components of this work owned by others than the author(s) must be honored. Abstracting with credit is permitted. To copy otherwise, or republish, to post on servers or to redistribute to lists, requires prior specific permission and/or a fee. Request permissions from permissions@acm.org.},
	shorttitle = {{FairSNA}},
	url = {https://dl.acm.org/doi/10.1145/3653711},
	doi = {10.1145/3653711},
	language = {EN},
	urldate = {2026-01-07},
	journal = {ACM Computing Surveys},
	author = {Saxena, Akrati and Fletcher, George and Pechenizkiy, Mykola},
	month = apr,
	year = {2024},
	note = {Publisher: ACMPUB27New York, NY},
}

@article{wilson_gender_2024,
	title = {Gender, {Race}, and {Intersectional} {Bias} in {Resume} {Screening} via {Language} {Model} {Retrieval}},
	volume = {7},
	copyright = {Copyright (c) 2024 Association for the Advancement of Artificial Intelligence},
	issn = {3065-8365},
	url = {https://ojs.aaai.org/index.php/AIES/article/view/31748},
	doi = {10.1609/aies.v7i1.31748},
	language = {en},
	number = {1},
	urldate = {2026-01-08},
	journal = {Proceedings of the AAAI/ACM Conference on AI, Ethics, and Society},
	author = {Wilson, Kyra and Caliskan, Aylin},
	month = oct,
	year = {2024},
	pages = {1578--1590},
}

@inproceedings{siddique_who_2024,
	address = {Miami, Florida, USA},
	title = {Who is better at math, {Jenny} or {Jingzhen}? {Uncovering} {Stereotypes} in {Large} {Language} {Models}},
	shorttitle = {Who is better at math, {Jenny} or {Jingzhen}?},
	url = {https://aclanthology.org/2024.emnlp-main.1035/},
	doi = {10.18653/v1/2024.emnlp-main.1035},
	urldate = {2026-01-08},
	booktitle = {Proceedings of the 2024 {Conference} on {Empirical} {Methods} in {Natural} {Language} {Processing}},
	publisher = {Association for Computational Linguistics},
	author = {Siddique, Zara and Turner, Liam and Espinosa-Anke, Luis},
	editor = {Al-Onaizan, Yaser and Bansal, Mohit and Chen, Yun-Nung},
	month = nov,
	year = {2024},
	pages = {18601--18619},
}

@inproceedings{weng_rolenet_2007,
	address = {New York, NY, USA},
	series = {{MIR} '07},
	title = {{RoleNet}: treat a movie as a small society},
	isbn = {978-1-59593-778-0},
	shorttitle = {{RoleNet}},
	url = {https://doi.org/10.1145/1290082.1290092},
	doi = {10.1145/1290082.1290092},
	urldate = {2025-12-31},
	booktitle = {Proceedings of the international workshop on {Workshop} on multimedia information retrieval},
	publisher = {Association for Computing Machinery},
	author = {Weng, Chung-Yi and Chu, Wei-Ta and Wu, Ja-Ling},
	month = sep,
	year = {2007},
	pages = {51--60},
}

@book{wasserman_social_1994, place={Cambridge}, series={Structural Analysis in the Social Sciences}, title={Social Network Analysis: Methods and Applications}, publisher={Cambridge University Press}, author={Wasserman, Stanley and Faust, Katherine}, year={1994}, collection={Structural Analysis in the Social Sciences}}

@inproceedings{proebsting_identity-related_2025,
	address = {New York, NY, USA},
	series = {{EAAMO} '25},
	title = {Identity-related {Speech} {Suppression} in {Generative} {AI} {Content} {Moderation}},
	isbn = {979-8-4007-2140-3},
	url = {https://doi.org/10.1145/3757887.3763010},
	doi = {10.1145/3757887.3763010},
	urldate = {2026-01-09},
	booktitle = {Proceedings of the 5th {ACM} {Conference} on {Equity} and {Access} in {Algorithms}, {Mechanisms}, and {Optimization}},
	publisher = {Association for Computing Machinery},
	author = {Proebsting, Grace and Anigboro, Oghenefejiro Isaacs and Crawford, Charlie M. and Metaxa, Danaé and Friedler, Sorelle A.},
	month = nov,
	year = {2025},
	pages = {185--217},
}

@inproceedings{liu_racial_2024,
	address = {New York, NY, USA},
	series = {{EAAMO} '24},
	title = {Racial {Steering} by {Large} {Language} {Models}: {A} {Prospective} {Audit} of {GPT}-4 on {Housing} {Recommendations}},
	isbn = {979-8-4007-1222-7},
	shorttitle = {Racial {Steering} by {Large} {Language} {Models}},
	url = {https://doi.org/10.1145/3689904.3694709},
	doi = {10.1145/3689904.3694709},
	urldate = {2026-01-09},
	booktitle = {Proceedings of the 4th {ACM} {Conference} on {Equity} and {Access} in {Algorithms}, {Mechanisms}, and {Optimization}},
	publisher = {Association for Computing Machinery},
	author = {Liu, Eric Justin and So, Wonyoung and Hosoi, Peko and D'Ignazio, Catherine},
	month = oct,
	year = {2024},
	pages = {1--13},
}

@inproceedings{yu_unpacking_2022,
	address = {Cham},
	title = {Unpacking {Gender} {Stereotypes} in {Film} {Dialogue}},
	isbn = {978-3-031-19097-1},
	doi = {10.1007/978-3-031-19097-1_26},
	language = {en},
	booktitle = {Social {Informatics}},
	publisher = {Springer International Publishing},
	author = {Yu, Yulin and Hao, Yucong and Dhillon, Paramveer},
	editor = {Hopfgartner, Frank and Jaidka, Kokil and Mayr, Philipp and Jose, Joemon and Breitsohl, Jan},
	year = {2022},
	pages = {398--405},
}

@inproceedings{gautam_stop_2024,
	address = {Bangkok, Thailand},
	title = {Stop! {In} the {Name} of {Flaws}: {Disentangling} {Personal} {Names} and {Sociodemographic} {Attributes} in {NLP}},
	shorttitle = {Stop! {In} the {Name} of {Flaws}},
	url = {https://aclanthology.org/2024.gebnlp-1.20/},
	doi = {10.18653/v1/2024.gebnlp-1.20},
	urldate = {2026-01-10},
	booktitle = {Proceedings of the 5th {Workshop} on {Gender} {Bias} in {Natural} {Language} {Processing} ({GeBNLP})},
	publisher = {Association for Computational Linguistics},
	author = {Gautam, Vagrant and Subramonian, Arjun and Lauscher, Anne and Keyes, Os},
	editor = {Faleńska, Agnieszka and Basta, Christine and Costa-jussà, Marta and Goldfarb-Tarrant, Seraphina and Nozza, Debora},
	month = aug,
	year = {2024},
	pages = {323--337},
}

@article{ugander_anatomy_2011,
  title={The Anatomy of the Facebook Social Graph},
  author={Johan Ugander and Brian Karrer and Lars Backstrom and Cameron A. Marlow},
  year={2011}
}

@article{mcpherson_birds_2001,
 ISSN = {03600572, 15452115},
 URL = {http://www.jstor.org/stable/2678628},
 author = {McPherson, Miller and Smith-Lovin, Lynn  and Cook, James M. },
 journal = {Annual Review of Sociology},
 pages = {415--444},
 publisher = {Annual Reviews},
 title = {Birds of a Feather: Homophily in Social Networks},
 urldate = {2026-01-10},
 volume = {27},
 year = {2001}
}

@article{kagan_using_2020,
	title = {Using data science to understand the film industry’s gender gap},
	volume = {6},
	copyright = {2020 The Author(s)},
	issn = {2055-1045},
	url = {https://www.nature.com/articles/s41599-020-0436-1},
	doi = {10.1057/s41599-020-0436-1},
	language = {en},
	number = {1},
	urldate = {2026-01-10},
	journal = {Palgrave Communications},
	author = {Kagan, Dima and Chesney, Thomas and Fire, Michael},
	month = may,
	year = {2020},
	note = {Publisher: Palgrave},
	pages = {92},
}

@article{kumar_gender_2022,
	title = {Gender {Stereotypes} in {Hollywood} {Movies} and {Their} {Evolution} over {Time}: {Insights} from {Network} {Analysis}},
	volume = {6},
	copyright = {http://creativecommons.org/licenses/by/3.0/},
	issn = {2504-2289},
	shorttitle = {Gender {Stereotypes} in {Hollywood} {Movies} and {Their} {Evolution} over {Time}},
	url = {https://www.mdpi.com/2504-2289/6/2/50},
	doi = {10.3390/bdcc6020050},
	language = {en},
	number = {2},
	urldate = {2026-01-10},
	journal = {Big Data and Cognitive Computing},
	author = {Kumar, Arjun M. and Goh, Jasmine Y. Q. and Tan, Tiffany H. H. and Siew, Cynthia S. Q.},
	month = jun,
	year = {2022},
	note = {Publisher: Multidisciplinary Digital Publishing Institute},
	pages = {50},
}

@inproceedings{sap_connotation_2017,
	address = {Copenhagen, Denmark},
	title = {Connotation {Frames} of {Power} and {Agency} in {Modern} {Films}},
	url = {https://aclanthology.org/D17-1247/},
	doi = {10.18653/v1/D17-1247},
	urldate = {2026-01-11},
	booktitle = {Proceedings of the 2017 {Conference} on {Empirical} {Methods} in {Natural} {Language} {Processing}},
	publisher = {Association for Computational Linguistics},
	author = {Sap, Maarten and Prasettio, Marcella Cindy and Holtzman, Ari and Rashkin, Hannah and Choi, Yejin},
	editor = {Palmer, Martha and Hwa, Rebecca and Riedel, Sebastian},
	month = sep,
	year = {2017},
	pages = {2329--2334},
}

@article{lv_storyrolenet_2018,
	title = {{StoryRoleNet}: {Social} {Network} {Construction} of {Role} {Relationship} in {Video}},
	volume = {6},
	issn = {2169-3536},
	shorttitle = {{StoryRoleNet}},
	url = {https://ieeexplore.ieee.org/document/8353310},
	doi = {10.1109/ACCESS.2018.2832087},
	urldate = {2026-01-11},
	journal = {IEEE Access},
	author = {Lv, Jinna and Wu, Bin and Zhou, Lili and Wang, Han},
	year = {2018},
	pages = {25958--25969},
}

@inproceedings{Xu_Ma_2025, address={Albuquerque, New Mexico}, title={LLM The Genius Paradox: A Linguistic and Math Expert’s Struggle with Simple Word-based Counting Problems}, ISBN={979-8-89176-189-6}, url={https://aclanthology.org/2025.naacl-long.172/}, DOI={10.18653/v1/2025.naacl-long.172}, abstractNote={Interestingly, LLMs yet struggle with some basic tasks that humans find trivial to handle, e.g., counting the number of character r’s in the word “strawberry”. There are several popular conjectures (e.g., tokenization, architecture and training data) regarding the reason for deficiency of LLMs in simple word-based counting problems, sharing the similar belief that such failure stems from model pretraining hence probably inevitable during deployment. In this paper, we carefully design multiple evaluation settings to investigate validity of prevalent conjectures. Meanwhile, we measure transferability of advanced mathematical and coding reasoning capabilities from specialized LLMs to simple counting tasks. Although specialized LLMs suffer from counting problems as well, we find conjectures about inherent deficiency of LLMs invalid and further seek opportunities to elicit knowledge and capabilities from LLMs which are beneficial to counting tasks. Compared with strategies such as finetuning and in-context learning that are commonly adopted to enhance performance on new or challenging tasks, we show that engaging reasoning is the most robust and efficient way to help LLMs better perceive tasks with more accurate responses.We hope our conjecture validation design could provide insights to study future critical failure modes of LLMs. Based on challenges in transferring advanced capabilities to much simpler tasks, we call for more attention to model capability acquisition and evaluation. We also highlight the importance of cultivating consciousness of “reasoning before responding” during model pretraining.}, booktitle={Proceedings of the 2025 Conference of the Nations of the Americas Chapter of the Association for Computational Linguistics: Human Language Technologies (Volume 1: Long Papers)}, publisher={Association for Computational Linguistics}, author={Xu, Nan and Ma, Xuezhe}, editor={Chiruzzo, Luis and Ritter, Alan and Wang, Lu}, year={2025}, month=apr, pages={3344–3370} }

@inproceedings{agarwal_parsing_2014,
	address = {Gothenburg, Sweden},
	title = {Parsing {Screenplays} for {Extracting} {Social} {Networks} from {Movies}},
	url = {https://aclanthology.org/W14-0907/},
	doi = {10.3115/v1/W14-0907},
	urldate = {2026-01-11},
	booktitle = {Proceedings of the 3rd {Workshop} on {Computational} {Linguistics} for {Literature} ({CLFL})},
	publisher = {Association for Computational Linguistics},
	author = {Agarwal, Apoorv and Balasubramanian, Sriramkumar and Zheng, Jiehan and Dash, Sarthak},
	editor = {Feldman, Anna and Kazantseva, Anna and Szpakowicz, Stan},
	month = apr,
	year = {2014},
	pages = {50--58},
}

@inproceedings{park_generative_2023,
	address = {New York, NY, USA},
	series = {{UIST} '23},
	title = {Generative {Agents}: {Interactive} {Simulacra} of {Human} {Behavior}},
	isbn = {979-8-4007-0132-0},
	shorttitle = {Generative {Agents}},
	url = {https://doi.org/10.1145/3586183.3606763},
	doi = {10.1145/3586183.3606763},
	urldate = {2026-01-11},
	booktitle = {Proceedings of the 36th {Annual} {ACM} {Symposium} on {User} {Interface} {Software} and {Technology}},
	publisher = {Association for Computing Machinery},
	author = {Park, Joon Sung and O'Brien, Joseph and Cai, Carrie Jun and Morris, Meredith Ringel and Liang, Percy and Bernstein, Michael S.},
	month = oct,
	year = {2023},
	pages = {1--22},
}

@misc{slumdog2008,
  title =   {Slumdog Millionaire},
  author = {Boyle, D. and Tandan, L.},
  year = {2008},
  publisher = {Fox Searchlight Pictures}
}

@article{Newman_2003, title={Mixing patterns in networks}, volume={67}, DOI={10.1103/PhysRevE.67.026126}, abstractNote={We study assortative mixing in networks, the tendency for vertices in networks to be connected to other vertices that are like (or unlike) them in some way. We consider mixing according to discrete characteristics such as language or race in social networks and scalar characteristics such as age. As a special example of the latter we consider mixing according to vertex degree, i.e., according to the number of connections vertices have to other vertices: do gregarious people tend to associate with other gregarious people? We propose a number of measures of assortative mixing appropriate to the various mixing types, and apply them to a variety of real-world networks, showing that assortative mixing is a pervasive phenomenon found in many networks. We also propose several models of assortatively mixed networks, both analytic ones based on generating function methods, and numerical ones based on Monte Carlo graph generation techniques. We use these models to probe the properties of networks as their level of assortativity is varied. In the particular case of mixing by degree, we find strong variation with assortativity in the connectivity of the network and in the resilience of the network to the removal of vertices.}, number={2}, journal={Physical Review E}, publisher={American Physical Society}, author={Newman, M. E. J.}, year={2003}, month=feb, pages={026126} }

@article{laniado_gender_2016,
	title = {Gender homophily in online dyadic and triadic relationships},
	volume = {5},
	issn = {2193-1127},
	url = {https://doi.org/10.1140/epjds/s13688-016-0080-6},
	doi = {10.1140/epjds/s13688-016-0080-6},
	language = {en},
	number = {1},
	urldate = {2026-01-13},
	journal = {EPJ Data Science},
	author = {Laniado, David and Volkovich, Yana and Kappler, Karolin and Kaltenbrunner, Andreas},
	month = may,
	year = {2016},
	pages = {19},
}

@inproceedings{Durak_Pinar_Kolda_Seshadhri_2012, address={New York, NY, USA}, series={CIKM ’12}, title={Degree relations of triangles in real-world networks and graph models}, ISBN={978-1-4503-1156-4}, url={https://dl.acm.org/doi/10.1145/2396761.2398503}, DOI={10.1145/2396761.2398503}, abstractNote={Triangles are an important building block and distinguishing feature of real-world networks, but their structure is still poorly understood. Despite numerous reports on the abundance of triangles, there is very little information on what these triangles look like. We initiate the study of degree-labeled triangles, - specifically, degree homogeneity versus heterogeneity in triangles. This yields new insight into the structure of real-world graphs. We observe that networks coming from social and collaborative situations are dominated by homogeneous triangles, i.e., degrees of vertices in a triangle are quite similar to each other. On the other hand, information networks (e.g., web graphs) are dominated by heterogeneous triangles, i.e., the degrees in triangles are quite disparate. Surprisingly, nodes within the top 1% of degrees participate in the vast majority of triangles in heterogeneous graphs. We investigate whether current graph models reproduce the types of triangles that are observed in real data and observe that most models fail to accurately capture these salient features.}, booktitle={Proceedings of the 21st ACM international conference on Information and knowledge management}, publisher={Association for Computing Machinery}, author={Durak, Nurcan and Pinar, Ali and Kolda, Tamara G. and Seshadhri, C.}, year={2012}, month=oct, pages={1712–1716}, collection={CIKM ’12} }

@article{lambrecht_algorithmic_2019,
  title={Algorithmic bias? An empirical study of apparent gender-based discrimination in the display of STEM career ads},
  author={Lambrecht, Anja and Tucker, Catherine},
  journal={Management science},
  volume={65},
  number={7},
  pages={2966--2981},
  year={2019},
  publisher={INFORMS}
}

@article{metaxa_search_2019,
	title = {Search {Media} and {Elections}: {A} {Longitudinal} {Investigation} of {Political} {Search} {Results}},
	volume = {3},
	shorttitle = {Search {Media} and {Elections}},
	url = {https://dl.acm.org/doi/10.1145/3359231},
	doi = {10.1145/3359231},
	number = {CSCW},
	urldate = {2026-01-13},
	journal = {Proc. ACM Hum.-Comput. Interact.},
	author = {Metaxa, Danaë and Park, Joon Sung and Landay, James A. and Hancock, Jeff},
	month = nov,
	year = {2019},
	pages = {129:1--129:17},
}

@article{wang_lower_2024,
	title = {Lower {Quantity}, {Higher} {Quality}: {Auditing} {News} {Content} and {User} {Perceptions} on {Twitter}/{X} {Algorithmic} versus {Chronological} {Timelines}},
	volume = {8},
	shorttitle = {Lower {Quantity}, {Higher} {Quality}},
	url = {https://doi.org/10.1145/3687046},
	doi = {10.1145/3687046},
	number = {CSCW2},
	urldate = {2026-01-10},
	journal = {Proc. ACM Hum.-Comput. Interact.},
	author = {Wang, Stephanie and Huang, Shengchun and Zhou, Alvin and Metaxa, Danaë},
	month = nov,
	year = {2024},
	pages = {507:1--507:25},
}

@inproceedings{bolukbasi_man_2016,
	title = {Man is to {Computer} {Programmer} as {Woman} is to {Homemaker}? {Debiasing} {Word} {Embeddings}},
	volume = {29},
	shorttitle = {Man is to {Computer} {Programmer} as {Woman} is to {Homemaker}?},
	url = {https://proceedings.neurips.cc/paper_files/paper/2016/hash/a486cd07e4ac3d270571622f4f316ec5-Abstract.html},
	urldate = {2026-01-13},
	booktitle = {Advances in {Neural} {Information} {Processing} {Systems}},
	publisher = {Curran Associates, Inc.},
	author = {Bolukbasi, Tolga and Chang, Kai-Wei and Zou, James Y and Saligrama, Venkatesh and Kalai, Adam T},
	year = {2016},
}

@article{caliskan_semantics_2017,
	title = {Semantics derived automatically from language corpora contain human-like biases},
	volume = {356},
	url = {https://www.science.org/doi/abs/10.1126/science.aal4230},
	doi = {10.1126/science.aal4230},
	number = {6334},
	urldate = {2026-01-13},
	journal = {Science},
	publisher = {American Association for the Advancement of Science},
	author = {Caliskan, Aylin and Bryson, Joanna J. and Narayanan, Arvind},
	month = apr,
	year = {2017},
	pages = {183--186},
}

@article{robertson_auditing_2018,
	title = {Auditing {Partisan} {Audience} {Bias} within {Google} {Search}},
	volume = {2},
	copyright = {https://www.acm.org/publications/policies/copyright\_policy\#Background},
	issn = {2573-0142},
	url = {https://dl.acm.org/doi/10.1145/3274417},
	doi = {10.1145/3274417},
	language = {en},
	number = {CSCW},
	urldate = {2025-09-23},
	journal = {Proceedings of the ACM on Human-Computer Interaction},
	publisher = {Association for Computing Machinery (ACM)},
	author = {Robertson, Ronald E. and Jiang, Shan and Joseph, Kenneth and Friedland, Lisa and Lazer, David and Wilson, Christo},
	month = nov,
	year = {2018},
	pages = {1--22},
}

@article{saleem_muslim_2019,
	title = {Muslim {Americans}’ {Responses} to {Social} {Identity} {Threats}: {Effects} of {Media} {Representations} and {Experiences} of {Discrimination}},
	volume = {22},
	url = {https://doi.org/10.1080/15213269.2017.1302345},
	doi = {10.1080/15213269.2017.1302345},
	number = {3},
	journal = {Media Psychology},
	publisher = {Routledge},
	author = {Saleem, Muniba and Ramasubramanian, Srividya},
	year = {2019},
	note = {\_eprint: https://doi.org/10.1080/15213269.2017.1302345},
	pages = {373--393},
}

@article{leavitt_frozen_2015,
	title = {“{Frozen} in {Time}”: {The} {Impact} of {Native} {American} {Media} {Representations} on {Identity} and {Self}-{Understanding}},
	volume = {71},
	url = {https://spssi.onlinelibrary.wiley.com/doi/abs/10.1111/josi.12095},
	doi = {https://doi.org/10.1111/josi.12095},
	number = {1},
	journal = {Journal of Social Issues},
	author = {Leavitt, Peter A. and Covarrubias, Rebecca and Perez, Yvonne A. and Fryberg, Stephanie A.},
	year = {2015},
	note = {\_eprint: https://spssi.onlinelibrary.wiley.com/doi/pdf/10.1111/josi.12095},
	pages = {39--53},
}

@inproceedings{Iso_Pezeshkpour_Bhutani_Hruschka_2025, address={Albuquerque, New Mexico}, title={Evaluating Bias in LLMs for Job-Resume Matching: Gender, Race, and Education}, ISBN={979-8-89176-194-0}, url={https://aclanthology.org/2025.naacl-industry.55/}, DOI={10.18653/v1/2025.naacl-industry.55}, abstractNote={Large Language Models (LLMs) offer the potential to automate hiring by matching job descriptions with candidate resumes, streamlining recruitment processes, and reducing operational costs. However, biases inherent in these models may lead to unfair hiring practices, reinforcing societal prejudices and undermining workplace diversity. This study examines the performance and fairness of LLMs in job-resume matching tasks within the English language and U.S. context. It evaluates how factors such as gender, race, and educational background influence model decisions, providing critical insights into the fairness and reliability of LLMs in HR applications.Our findings indicate that while recent models have reduced biases related to explicit attributes like gender and race, implicit biases concerning educational background remain significant. These results highlight the need for ongoing evaluation and the development of advanced bias mitigation strategies to ensure equitable hiring practices when using LLMs in industry settings.}, booktitle={Proceedings of the 2025 Conference of the Nations of the Americas Chapter of the Association for Computational Linguistics: Human Language Technologies (Volume 3: Industry Track)}, publisher={Association for Computational Linguistics}, author={Iso, Hayate and Pezeshkpour, Pouya and Bhutani, Nikita and Hruschka, Estevam}, editor={Chen, Weizhu and Yang, Yi and Kachuee, Mohammad and Fu, Xue-Yong}, year={2025}, month=apr, pages={672–683} }

@inproceedings{Cyberey_Ji_Evans_2025, address={Suzhou, China}, title={Unsupervised Concept Vector Extraction for Bias Control in LLMs}, ISBN={979-8-89176-332-6}, url={https://aclanthology.org/2025.emnlp-main.1439/}, DOI={10.18653/v1/2025.emnlp-main.1439}, abstractNote={Large language models (LLMs) are known to perpetuate stereotypes and exhibit biases. Various strategies have been proposed to mitigate these biases, but most work studies biases as a black-box problem without considering how concepts are represented within the model. We adapt techniques from representation engineering to study how the concept of “gender” is represented within LLMs. We introduce a new method that extracts concept representations via probability weighting without labeled data and efficiently selects a steering vector for measuring and manipulating the model’s representation. We develop a projection-based method that enables precise steering of model predictions and demonstrate its effectiveness in mitigating gender bias in LLMs and show that it also generalizes to racial bias.}, booktitle={Proceedings of the 2025 Conference on Empirical Methods in Natural Language Processing}, publisher={Association for Computational Linguistics}, author={Cyberey, Hannah and Ji, Yangfeng and Evans, David}, editor={Christodoulopoulos, Christos and Chakraborty, Tanmoy and Rose, Carolyn and Peng, Violet}, year={2025}, month=nov, pages={28333–28355} }

@inproceedings{tseng_ownership_2025,
author = {Tseng, Emily and Young, Meg and Le Qu\'{e}r\'{e}, Marianne Aubin and Rinehart, Aimee and Suresh, Harini},
title = {"Ownership, Not Just Happy Talk": Co-Designing a Participatory Large Language Model for Journalism},
year = {2025},
isbn = {9798400714825},
publisher = {Association for Computing Machinery},
address = {New York, NY, USA},
url = {https://doi.org/10.1145/3715275.3732198},
doi = {10.1145/3715275.3732198},
booktitle = {Proceedings of the 2025 ACM Conference on Fairness, Accountability, and Transparency},
pages = {3119–3130},
numpages = {12},
location = {
},
series = {FAccT '25}
}

@article{Valentowitsch_2023, title={Hollywood caught in two worlds? The impact of the Bechdel test on the international box office performance of cinematic films}, volume={34}, ISSN={1573-059X}, DOI={10.1007/s11002-022-09652-5}, abstractNote={The Bechdel test is increasingly used in academia as a quality indicator for the portrayal of women in films. Previous studies explored how passing the Bechdel test affects box office earnings. However, earlier considerations were all limited to the US market. Therefore, the impact of the Bechdel test on international box office receipts is still unclear. To fill this research gap, this study examines the box office effect for internationally released films at the country level. Using a sample of 515 randomly selected Hollywood films, it is shown that passing the test significantly improves international box office earnings. However, the results also show that the effect depends on the level of socioeconomic development in the respective countries. Cultural discount theory is used to explain the empirical findings.}, number={2}, journal={Marketing Letters}, author={Valentowitsch, Johann}, year={2023}, month=june, pages={293–308}, language={en} }

@inproceedings{gorinski_movie_2015,
	address = {Denver, Colorado},
	title = {Movie {Script} {Summarization} as {Graph}-based {Scene} {Extraction}},
	url = {https://aclanthology.org/N15-1113/},
	doi = {10.3115/v1/N15-1113},
	urldate = {2026-03-12},
	booktitle = {Proceedings of the 2015 {Conference} of the {North} {American} {Chapter} of the {Association} for {Computational} {Linguistics}: {Human} {Language} {Technologies}},
	publisher = {Association for Computational Linguistics},
	author = {Gorinski, Philip John and Lapata, Mirella},
	editor = {Mihalcea, Rada and Chai, Joyce and Sarkar, Anoop},
	month = may,
	year = {2015},
	pages = {1066--1076},
}

@article{Park_Oh_Jo_2012, title={Social network analysis in a movie using character-net}, volume={59}, ISSN={1380-7501}, DOI={10.1007/s11042-011-0725-1}, abstractNote={There have been various approaches to analyzing movie stories using social networks. Social network analysis is an effective means to extract semantic information from movies. Movie analysis through social relationships among characters can support various types of information retrieval better than audio-visual feature analysis. The relationships among characters form the main structure of the story. Therefore, through social network analysis among characters, movie story information such as the major roles and the corresponding communities can be determined. Progression of most movie stories is done by characters, and the scriptwriter or director narrates the story and relationships among characters using character dialogs. A dialog has a direction and time that supplies information. Therefore, the dialog is better for constructing social networks of characters than the co-appearance. Additionally, through social networks using the dialog, we can extract accurate movie stories such as classification of major, minor or extra roles, community clustering, and sequence detection. To achieve this, we propose a Character-net that can represent the relationships between characters using dialogs, and a method that can extract the sequences via clustering communities composed of characters. Our experiments show that our proposed method can efficiently detect sequences.}, number={2}, journal={Multimedia Tools Appl.}, author={Park, Seung-Bo and Oh, Kyeong-Jin and Jo, Geun-Sik}, year={2012}, month=july, pages={601–627} }

\appendix

\section{Screenplay generation prompts}
Two prompts used to generate screenplays using LLMs in a two-step process. 

\begin{figure}[H]
\framebox{
    \begin{minipage}{\linewidth}
    You are a helpful AI assistant.
    I will provide a movie's synopsis where all character names are redacted as PERSON IDs (e.g., PERSON 1, PERSON 33) and the number of scenes in the movie.

    Your tasks:
    \begin{enumerate}
        \item Assign each PERSON ID a unique, consistent character name based on the roles and relationships implied in the synopsis.
        \item Generate a numbered list of scenes for the movie based on the synopsis. Each scene must have:
        \begin{itemize}
            \item "number": integer
            \item "description": one concise, single-sentence summary of events, including the names of all characters present in that scene
            \item "characters": list of character names present in the scene
        \end{itemize}
    \end{enumerate}
    Movie Synopsis:
    \{synopsis\}

    Number of Scenes:
    \{num\_scenes\}
    \end{minipage}
}
\caption{Prompt used to generate structured movie scenes from anonymized synopses.}
\label{fig:screenplay_prompt}
\end{figure}

\begin{figure}[H]
\framebox{
    \begin{minipage}{\linewidth}
    You are a helpful AI assistant.
    I will provide a movie's synopsis and a summary of a scene from the movie. The movie's synopsis have character names redacted as PERSON IDs (e.g., PERSON 1, PERSON 33). The scene description provides the real character names. Use the real character names in the screenplay. Based on the movie synopsis and scene description, please write a detailed screenplay for the scene.

    Requirements:
    \begin{itemize}
        \item Only character names should be in ALL CAPS before their dialogue.
        \item Do NOT use ALL CAPS for scene descriptions, stage directions, or any other text.
        \item Use professional screenplay formatting.
    \end{itemize}
    
    Movie Synopsis:
    \{synopsis\}

    Scene Description:
    SCENE \{scene.number\}: \{scene.description\}
    \end{minipage}
    }
\caption{Prompt used to generate structured screenplays from a given movie scene and anonymized synopsis.}
\label{fig:scene_prompt}
\end{figure}

\section{SNA features used in Bechdel test model}
We record the list of 43 SNA features used to predict whether a script passed the third part of the Bechdel test, in keeping with the method used by \citet{agarwal_key_2015}.

\begin{table}[H]
    \centering
    \caption{List of 43 SNA features used to determine whether movie script passed Test 3 of the Bechdel test}
    \small
     \begin{tabularx}{\textwidth}{|c|X|c|}
     \hline
     \# & Feature & Type \\
     \hline
     1--4 & Degree centrality (Min, Max, Mean, SD) & Normalized\\ 
     5--8 & Closeness centrality (Min, Max, Mean, SD) & Normalized \\ 
     9--12 & Betweenness centrality (Min, Max, Mean, SD) & Normalized\\ 
     13--16 & Number of male characters a female character is connected to (Min, Max, Mean, SD)& Absolute\\
     17--20 & Number of female characters a female character is connected to (Min, Max, Mean, SD) & Absolute\\
     21--24 & Number of male characters in common between 2 female characters (Min, Max, Mean, SD) & Absolute \\
     25--28 & Number of female characters in common between 2 female characters (Min, Max, Mean, SD) & Absolute\\
     29 & Ratio of number of female characters to male characters & Normalized\\ 
     30 & Ratio of number of female characters to total characters & Normalized\\
     31 & Percentage of female characters that form a 3-clique with another male character and female character & Normalized\\
     32--34 & Percentage of female characters in the in top 5 main characters (based on each of the 3 centrality measures) & Normalized\\ 
     35--37 & Boolean recording whether the main character is female (based on each of the 3 centrality measures) & Normalized\\
     38--40 & Boolean recording whether any female character connects another female character to the main male character (based on each of the 3 centrality measures) & Normalized\\
     41--43 & Percentage of female characters that connect the main male character to another female character (based on each of the 3 centrality measures & Normalized \\
     \hline
     \end{tabularx}
        \label{fig:43 SNA features}
\end{table}

\section{Bechdel test model validation}

\begin{table}[H]
 \centering
  \caption{Precision, recall, and F1 scores for our Bechdel test model and~\citet{agarwal_key_2015} on human-written screenplays, evaluated against human annotations from the Bechdel Test Movie List as ground truth. Our model performs comparably to~\citet{agarwal_key_2015}.}
     \begin{tabular}{|l|ccc|ccc|}
     \hline
     \textbf{Bechdel Test Criteria} & \multicolumn{3}{c|}{\textbf{Our Model}} & \multicolumn{3}{c|}{\textbf{Agarwal et al.}} \\
      & P & R & F1 & P & R & F1\\
     \hline
     Pass Test 1 & 0.96 & 0.97 & \textbf{0.97} & 0.97 & 0.96 & 0.96\\
     Pass Test 2 & 0.77 & 0.78 & \textbf{0.78} & 0.67 & 0.90 & 0.77\\
     Pass Test 3 & 0.65 & 0.80 & 0.72 & 0.83 & 0.66 & \textbf{0.73} \\
     \hline
     \end{tabular}
 \label{tab:bechdel-validation}
 \end{table}

 \begin{table}[H]
 \centering
 \caption{Precision, recall, and F1 scores for Test 3 (i.e., overall pass/fail) of our Bechdel test model on human-written and LLM-generated screenplays, using human annotations from the Bechdel Test Movie List for human-written screenplays and our own hand-annotations of 100 screenplays per model for LLM-generated screenplays as ground truth. Our model performs comparably across both screenplay types.}
     \begin{tabular}{|l|ccc|}
     \hline
     \textbf{Script Type} & \multicolumn{3}{c|}{\textbf{Our Model}} \\
      & P & R & F1\\
     \hline
     Human & 0.65 & 0.80 & 0.72 \\
     GPT & 0.68 & 0.96 & \textbf{0.80} \\
     Gemini & 0.56 & \textbf{0.97} & 0.71 \\
     Claude & \textbf{0.69} & 0.92 & 0.79 \\
     \hline
     \end{tabular}
 \label{tab:llm-validation}
 \end{table} 

\section{List of films}
\begin{center}
\scriptsize
\setlength{\tabcolsep}{4pt}

{
\begin{longtable}{llll}
\caption{All 783 films in the dataset before network conversion, listed alphabetically.}\\
\label{tab:films-full}\\

\toprule
Film (year) &  &  &  \\
\midrule
\endfirsthead

\multicolumn{4}{c}{\tablename\ \thetable\ -- \textit{continued}} \\
\toprule
Film (year) &  &  &  \\
\midrule
\endhead

\midrule
\multicolumn{4}{r}{\textit{continued on next page}} \\
\endfoot

\bottomrule
\endlastfoot

\textit{9} (2009) &
\textit{12} (2003) &
\textit{42} (2019) &
\textit{2012} (2009) \\
\textit{12 Years a Slave} (2013) &
\textit{127 Hours} (2010) &
\textit{17 Again} (2009) &
\textit{20th Century} (2016) \\
\textit{28 Days Later} (2002) &
\textit{30 Minutes or Less} (2001) &
\textit{44 Inch Chest} (2009) &
\textit{48 Hrs.} (1982) \\
\textit{8MM} (1999) &
\textit{A Few Good Men} (1992) &
\textit{A Most Violent Year} (2014) &
\textit{A Nightmare on Elm Street} (2010) \\
\textit{A Perfect World} (1993) &
\textit{A Quiet Place} (2016) &
\textit{A Real Pain} (2024) &
\textit{A Scanner Darkly} (2006) \\
\textit{A Serious Man} (2009) &
\textit{A Walk to Remember} (2002) &
\textit{Ad Astra} (2016) &
\textit{Adaptation} (2022) \\
\textit{After.Life} (2009) &
\textit{Agnes of God} (1985) & 
\textit{Air} (2004) &
\textit{Air Force One} (2003) \\
\textit{Airplane} (1980) &
\textit{Aladdin} (1992) & 
\textit{Ali} (2001) &
\textit{Alien} (2025) \\
\textit{Alien 3} (1992) &
\textit{Alien Nation} (1988) &
\textit{Alien v. Predator} (2004) &
\textit{Aliens} (2014) \\
\textit{All About Eve} (1950)&
\textit{All About Steve} (2009) &
\textit{All of Us Strangers} (2023) &
\textit{Almost Famous} (2000) \\
\textit{Amadeus} (1984) &
\textit{Amelia} (2003) &
\textit{American Beauty} (1945) & 
\textit{American Gangster} (2007) \\
\textit{American Graffiti} (1973) &
\textit{American History X} (1998) &
\textit{American Hustle} (2013) &
\textit{American Outlaws} (2023) \\
\textit{American Pie} (1999) &
\textit{American Sniper} (2014) & 
\textit{American Werewolf in London} (1981) & 
\textit{Amour} (2023) \\
\textit{An Education} (2009) &
\textit{Analyze This} (1999) &
\textit{Anastasia} (1997) &
\textit{Annie Hall} (1977) \\
\textit{Anora} (2024) & 
\textit{Antitrust} (2001) &
\textit{Antz} (1998) & 
\textit{Apocalypse Now} (1979) \\
\textit{Apt Pupil} (1998) &
\textit{Arbitrage} (2012) &
\textit{Arctic Blue} (1993) &
\textit{Argo} (2017) \\
\textit{Armageddon} (1998) &
\textit{Army of Darkness} (1992) &
\textit{Arsenic and Old Lace} (1969) &
\textit{As Good As It Gets} (1997) \\
\textit{Assassins} (1995) &
\textit{Asteroid City} (2023) &
\textit{Autumn in New York} (2000) &
\textit{Avatar} (2009) \\
\textit{Babel} (2019) &
\textit{Bachelor Party} (2009) &
\textit{Backdraft} (1991) &
\textit{Bad Boys} (2014) \\
\textit{Bad Day at Black Rock} (1955)&
\textit{Bad Lieutenant} (1992) &
\textit{Bad Santa} (2003) & 
\textit{Bad Teacher} (2011) \\
\textit{Barbie} (2023) &
\textit{Barry Lyndon} (1975) &
\textit{Barton Fink} (1991) & 
\textit{Basic} (2020) \\
\textit{Basic Instinct} (1992) &
\textit{Basquiat} (1996) &
\textit{Batman} (1943) & 
\textit{Bean} (2017) \\
\textit{Beasts of No Nation} (2015) &
\textit{Beasts of the Southern Wild} (2012) &
\textit{Beauty and the Beast} (1997) &
\textit{Beginners} (2018) \\
\textit{Being John Malkovich} (1999) &
\textit{Being the Ricardos} (2021) &
\textit{Being There} (2017) &
\textit{Belle} (2019) \\
\textit{Beloved} (1976) &
\textit{Big} (1988) &
\textit{Big Eyes} (2014) & 
\textit{Big Fish} (2004) \\
\textit{Black Rain} (2021) &
\textit{Blackberry} (2023) &
\textit{BlacKkKlansman} (2018) & 
\textit{Blade} (1973) \\
\textit{Blade II} (2002) &
\textit{Blade Runner} (1982) & 
\textit{Blood Simple} (1985) & 
\textit{Blue Valentine} (2010) \\
\textit{Blue Velvet} (1986) &
\textit{Body Heat} (1981) &
\textit{Body of Evidence} (1988) & 
\textit{Bones} (2016) \\
\textit{Bonfire of the Vanities} (1990) &
\textit{Bonnie and Clyde} (1967) & 
\textit{Boogie Nights} (1997) &
\textit{Bookworm} (2024) \\
\textit{Bottle Rocket} (2025) &
\textit{Bound} (2017) &
\textit{Braveheart} (1925) & 
\textit{Break} (1986) \\
\textit{Brick} (2006) &
\textit{Bridesmaids} (1989) & 
\textit{Broadcast News} (1987) & 
\textit{Broken Arrow} (2022) \\
\textit{Broken Embraces} (2009) &
\textit{Bruce Almighty} (2003) &
\textit{Buffy the Vampire Slayer} (1992) &
\textit{Bull Durham} (1988) \\
\textit{Buried} (2024) &
\textit{Burn After Reading} (2008) & 
\textit{Cable Guy} (1996) &
\textit{Capote} (2005) \\
\textit{Carrie} (1952) &
\textit{Cars 2} (2011) &
\textit{Case 39} (2009) &
\textit{Casino} (2011) \\
\textit{Cast Away} (2017) &
\textit{Catch Me If You Can} (1998) &
\textit{Cecil B. Demented} (2000) &
\textit{Cedar Rapids} (2011) \\
\textit{Cellular} (2020) &
\textit{Changeling} (2019) &
\textit{Chaos} (2021) &
\textit{Charade} (1963) \\
\textit{Chasing Amy} (1997) &
\textit{Cherry Falls} (2000) &
\textit{Chinatown} (1974) &
\textit{Boogie Nights} (1997)\\
\textit{Cinema Paradiso} (1998) &
\textit{Citizen Kane} (1941) & 
\textit{Clash of the Titans} (1981) &
\textit{Clerks} (1994) \\
\textit{Cliffhanger} (2026) &
\textit{Clueless} (1995) &
\textit{Cobb} (1994) &
\textit{Coco} (2000) \\
\textit{Cold Mountain} (2003) &
\textit{Colombiana} (2011) & 
\textit{Color of Night} (1994) &
\textit{Confessions of a Dangerous Mind} (2002) \\
\textit{Constantine} (2005) &
\textit{Copycat} (2024) &
\textit{Coraline} (2009) &
\textit{Coriolanus} (1997) \\
\textit{Corpse Bride} (2005) &
\textit{Cradle 2 the Grave} (2003) &
\textit{Crank} (2006) &
\textit{Crazy, Stupid, Love} (2011) \\
\textit{Creation} (2020) &
\textit{Crouching Tiger, Hidden Dragon} (2000) &
\textit{Cruel Intentions} (2016) &
\textit{Crying Game} (1992) \\
\textit{Cube} (2019) &
\textit{Custody} (2005) &
\textit{Dallas Buyers Club} (2013) &
\textit{Dances with Wolves} (1990) \\
\textit{Dark City} (1990) &
\textit{Dark Star} (1974) &
\textit{Darkman} (1992) & 
\textit{Date Night} (2024) \\
\textit{Dawn of the Dead} (2004) &
\textit{Day of the Dead} (1985) &
\textit{Days of Heaven} (1978) &
\textit{Dead Poets Society} (1989) \\
\textit{Deadpool} (2016) &
\textit{Dear White People} (2014) &
\textit{Death at a Funeral} (2007) &
\textit{Death to Smoochy} (2002) \\
\textit{Deception} (2011) &
\textit{Defiance} (1993) &
\textit{Despicable Me} (2013) &
\textit{Devil in a Blue Dress} (1995) \\
\textit{Die Hard} (1988) &
\textit{Die Hard 2} (1990) &
\textit{Diner} (1983) & 
\textit{Disturbia} (2007) \\
\textit{Django Unchained} (2012) &
\textit{Do the Right Thing} (1989) &
\textit{Dog Day Afternoon} (1975) & 
\textit{Dogma} (1999) \\
\textit{Donnie Brasco} (1997) &
\textit{Double Indemnity} (1973) & 
\textit{Drag Me to Hell} (2009) &
\textit{Dragonslayer} (2011) \\
\textit{Drive} (2022) &
\textit{Drop Dead Gorgeous} (2010) &
\textit{Duck Soup} (1942) & 
\textit{Dumb and Dumber} (1994) \\
\textit{Dune} (2021) & 
\textit{E.T.} (1982) &
\textit{Eagle Eye} (2008) &
\textit{Eastern Promises} (2007) \\
\textit{Easy A} (2010) & 
\textit{Ed Wood} (1994) &
\textit{Edward Scissorhands} (1990) & 
\textit{Election} (1999) \\
\textit{Elemental} (2023) &
\textit{Evlis} (2005) &
\textit{Enemy of the State} (1998) &
\textit{Enough} (2021) \\
\textit{Erin Brockovich} (2000) &
\textit{Escape From L.A.} (1996) &
\textit{Escape From New York} (1981) & 
\textit{Eternal Sunshine of the Spotless Mind} (2004) \\
\textit{Even Cowgirls Get the Blues} (1994) &
\textit{Event Horizon} (2017) &
\textit{Evil Dead} (2013) &
\textit{Ex Machina} (2015) \\
\textit{Excalibur} (1981) &
\textit{Extract} (2009) &
\textit{Fair Game} (2017) &
\textit{Fantastic Beasts and Where to Find Them} (2016) \\
\textit{Fantastic Four} (2015) &
\textit{Fargo} (1952) &
\textit{Fast Times at Ridgemont High} (1982) & 
\textit{Fatal Instinct} (2014) \\
\textit{Fear and Loathing in Las Vegas} (1998) & 
\textit{Feast} (2023) &
\textit{Ferrari} (2023) & 
\textit{Field of Dreams} (1989) \\
\textit{Fight Club} (1999) &
\textit{Final Destination} (2000) &
\textit{Final Destination 2} (2003) &
\textit{Finding Nemo} (2003) \\
\textit{Five Easy Pieces} (1995) &
\textit{Flash Gordon} (1980) &
\textit{Fletch} (1985) &
\textit{Flight} (1997) \\
\textit{Forrest Gump} (1994) &
\textit{Four Rooms} (1995) &
\textit{Foxcatcher} (2014) &
\textit{Fracture} (2020) \\
\textit{Fred Claus} (2007) &
\textit{Freddy vs. Jason} (2003) &
\textit{Friday the 13th} (2016) &
\textit{Fright Night} (2011) \\
\textit{From Dusk Till Dawn} (1996) &
\textit{From Here to Eternity} (2014) &
\textit{Frozen River} (1929) &
\textit{Fruitvale Station} (2013) \\
\textit{Funny People} (1981) & 
\textit{G.I. Jane} (1951) &
\textit{Gamer} (2024) &
\textit{Gandhi} (1982) \\
\textit{Gangs of New York} (1938) &
\textit{Garden State} (2024) &
\textit{Gattaca} (1997) & 
\textit{Get Carter} (1971) \\
\textit{Get Low} (2010) &
\textit{Get on Up} (2014) &
\textit{Get Out} (2017) & 
\textit{Ghost Ship} (2002) \\
\textit{Ghost World} (2001) &
\textit{Ghostbusters} (2016) &
\textit{Gladiator} (1992) &
\textit{Godzilla} (1998) \\
\textit{Good Will Hunting} (1997) &
\textit{Gothika} (2003) &
\textit{Grabbers} (2012) &
\textit{Gran Torino} (2008) \\
\textit{Grand Hotel} (2018) &
\textit{Gravity} (2014) &
\textit{Gremlins} (1984) & 
\textit{Gremlins 2} (1990) \\
\textit{Groundhog Day} (1993) &
\textit{Hackers} (1995) &
\textit{Hall Pass} (2011) &
\textit{Hancock} (1991) \\
\textit{Hanna} (2011) &
\textit{Hannah and Her Sisters} (1986) & 
\textit{Hannibal} (1972) & 
\textit{Happy Feet} (1991) \\
\textit{Hard Rain} (1998) &
\textit{Harold and Kumar Go to White Castle} (2004) &
\textit{Heathers} (1988) & 
\textit{Heavenly Creatures} (1994) \\
\textit{Heavy Metal} (1979) &
\textit{Hellboy} (2019) &
\textit{Hellraiser} (2022) &
\textit{Her} (Not Avail.) \\
\textit{Hesher} (2010) & 
\textit{High Fidelity} (2000) &
\textit{Highlander} (1986) &
\textit{His Girl Friday} (1940) \\
\textit{Hitchcock} (2012) &
\textit{Hollow Man} (2000) &
\textit{Horrible Bosses} (2011) &
\textit{Hot Tub Time Machine} (2010) \\
\textit{Hotel Rwanda} (2004) &
\textit{House of 1000 Corpses} (2003) &
\textit{How to Train Your Dragon} (2010) &
\textit{Hudson Hawk} (1991) \\
\textit{Human Nature} (2025) &
\textit{I Am Number Four} (2011) &
\textit{I am Sam} (2001) & 
\textit{I Love You Phillip Morris} (2010) \\
\textit{I Spit on Your Grave} (2010) &
\textit{I Still Know What You Did Last Summer} (1998) &
\textit{I, Robot} (2004) & 
\textit{In the Bedroom} (2024) \\
\textit{In the Loop} (2009) &
\textit{Inception} (2010) &
\textit{Boogie Nights} (1997) &
\textit{Indiana Jones and the Last Crusade} (1989) \\
\textit{Indiana Jones and the Temple of Doom} (1984) & 
\textit{Inglourious Bastards} (2009) &
\textit{Initiation} (2011) &
\textit{Insidious} (2008) \\
\textit{Insomnia} (2002) &
\textit{Interstellar} (2014) &
\textit{Interview with the Vampire} (1994) &
\textit{Into the Wild} (2007) \\
\textit{Into the Woods} (2014) &
\textit{Intolerable Cruelty} (2003) &
\textit{Invictus} (2009) &
\textit{It} (2017) \\
\textit{It Happened One Night} (1934) &
\textit{Jackie Brown} (1997) &
\textit{Jane Eyre} (1910) & 
\textit{Jason X} (2001) \\
\textit{Jaws 2} (1978) &
\textit{Jay and Silent Bob Strike Back} (2001) &
\textit{Jerry Maguire} (1996) & 
\textit{JFK} (2013) \\
\textit{John Wick} (2014) &
\textit{Jojo Rabbit} (2019) &
\textit{Judge Dredd} (1995) &
\textit{Juno} (2007) \\
\textit{Jurassic Park} (1993) & 
\textit{Jurassic Park III} (2001) &
\textit{Kids} (2019) &
\textit{Kill Your Darlings} (2024) \\
\textit{Killers of the Flower Moon} (2023) &
\textit{King Kong} (2012) &
\textit{Klute} (1971) &
\textit{Kill Bill} (2003) \\
\textit{King of Comedy} (1982) & 
\textit{Klute} (1971) &
\textit{Knocked Up} (2007) &
\textit{Kung Fu Panda} (2008) \\
\textit{La La Land} (2016) &
\textit{Lake Placid} (1999) &
\textit{Last Chance Harvey} (2008) &
\textit{Last Tango in Paris} (1972) \\
\textit{Law Abiding Citizens} (2021) &
\textit{Leaving Las Vegas} (1995) &
\textit{Legally Blonde} (2003) &
\textit{Legend} (2015) \\
\textit{Legion} (2024) &
\textit{Les Miserables} (2012) &
\textit{Liar Liar} (1997) &
\textit{Life} (1999) \\
\textit{Life As A House} (2001) & 
\textit{Life of Pi} (2012) &
\textit{Limitless} (2021) &
\textit{Lincoln} (2006) \\
\textit{Little Nicky} (2000) & 
\textit{Lock, Stock and Two Smoking Barrels} (1998) &
\textit{Logan} (2013) &
\textit{Looper} (2012) \\
\textit{Lord of War} (2005) & 
\textit{Lost Highway} (1997) &
\textit{Lost Horizon} (1937) &
\textit{Lost in Space} (1998) \\
\textit{Lost in Translation} (2003) &
\textit{Machete} (2010) &
\textit{Made} (2010) &
\textit{Magnolia} (1999) \\
\textit{Major League} (1989) &
\textit{Malcolm X} (1992) &
\textit{Malignant} (2021) &
\textit{Man in the Iron Mask} (1998) \\
\textit{Man on the Moon} (1999) &
\textit{Manhattan Murder Mystery} (1993) &
\textit{Manhunter} (1986) & 
\textit{Margaret} (2011) \\
\textit{Margin Call} (2011) &
\textit{Margot at the Wedding} (2007) &
\textit{Martha Marcy May Marlene} (2011) &
\textit{Marty} (2018) \\
\textit{Mary Poppins} (1964) & 
\textit{Max Payne} (2008) &
\textit{May December} (2023) &
\textit{Mean Streets} (1973) \\
\textit{Meet Joe Black} (1998) &
\textit{Meet John Doe} (1941) &
\textit{Megamind} (2010) &
\textit{Memento} (2000) \\
\textit{Memory} (2023) &
\textit{Men in Black} (1997) &
\textit{Metro} (2019) &
\textit{Miami Vice} (2006) \\
\textit{Midnight Cowboy} (1969) & 
\textit{Midnight Express} (1978) &
\textit{Midnight in Paris} (2011) &
\textit{Mighty Joe Young} (1998) \\
\textit{Milk} (2009) &
\textit{Minority Report} (2002) & 
\textit{Mirrors} (2008) &
\textit{Misery} (1990) \\
\textit{Mission Impossible} (1996) & 
\textit{Mission to Mars} (2000) &
\textit{Moneyball} (2011) &
\textit{Monkeybone} (2001) \\
\textit{Monte Carlo} (2011) &
\textit{Moon} (2009) &
\textit{Moonrise Kingdom} (2012) &
\textit{Moonstruck} (1987) \\
\textit{Mrs. Brown} (1997) &
\textit{Mud} (2013) &
\textit{Mulan} (1998) &
\textit{Music of the Heart} (1999) \\
\textit{My Girl} (1991) &
\textit{My Week with Marilyn} (2011) &
\textit{Mystery Men} (1999) &
\textit{Nashville} (1975) \\
\textit{Natural Born Killers} (1994) & 
\textit{Never Been Kissed} (1999) &
\textit{New York Minute} (2003) &
\textit{Newsies} (1992) \\
\textit{Next} (2007) &
\textit{Next Friday} (2000) &
\textit{Nick of Time} (1995) &
\textit{Nightbreed} (1990) \\
\textit{Nine} (2009) &
\textit{Ninja Assassin} (2009) &
\textit{No Country for Old Men} (2007) & 
\textit{No Hard Feelings} (2023) \\
\textit{Notting Hill} (1999) &
\textit{Oblivion} (2013) &
\textit{Ocean's Twelve} (2004) &
\textit{Office Space} (1999) \\
\textit{Only God Forgives} (2013) &
\textit{Onward} (2020) &
\textit{Oppenheimer} (2023) &
\textit{Ordinary People} (1980) \\
\textit{Origin} (2023) & 
\textit{Orphan} (2009) &
\textit{Pandorum} (2009) & 
\textit{Panic Room} (2002) \\
\textit{ParaNorman} (2012) & 
\textit{Pariah} (2011) & 
\textit{Passengers} (2016) & 
\textit{Paul} (2011) \\ 
\textit{Pearl Harbor} (2001) & 
\textit{Peeping Tom} (1960) & 
\textit{Peggy Sue Got Married} (1986) & 
\textit{Pet Sematary} (1989) \\ 
\textit{Philadelphia} (1993) & 
\textit{Phone Booth} (2003) &
\textit{Pi} (1998) & 
\textit{Pineapple Express} (2008) \\
\textit{Pirates of the Caribbean} (2003) & 
\textit{Pitch Black} (2000) &
\textit{Platoon} (1986) & 
\textit{Point Break} (1991) \\
\textit{Poor Things} (2023) & 
\textit{Precious} (2009) & 
\textit{Predator} (1987) & 
\textit{Pride and Prejudice} (2005) \\
\textit{Priest} (2011) & 
\textit{Priscilla} (2023) &
\textit{Prom Night} (1980) & 
\textit{Prometheus} (2002) \\
\textit{Psycho} (1960) & 
\textit{Public Enemies} (2009) & 
\textit{Pulp Fiction} (1994) & 
\textit{Purple Rain} (1984) \\ 
\textit{Queen of the Damned} (2002) &
\textit{Quiz Lady} (2023) &
\textit{Rachel Getting Married}  (2008) & 
\textit{Raging Bull} (1980) \\
\textit{Raising Arizona} (1987) & 
\textit{Real Genius} (1985) & 
\textit{Rear Window} (1954) &
\textit{Rebel Without A Cause} (1955) \\
 \textit{Red Planet} (2000) &
\textit{Red Riding Hood} (2011) &
\textit{Remember Me} (1981) &
\textit{Repo Man} (1984) \\
\textit{Resident Evil} (2002) &
\textit{Revolutionary Road} (2008) &
\textit{Ringu} (1998) & 
\textit{Rise of the Guardians} (2012) \\
\textit{Rise of the Planet of the Apes} (2011) & 
\textit{RocknRolla} (2008) &
\textit{Rocky} (1976) & 
\textit{Ronin} (1998) \\
\textit{Room} (2016) &
\textit{Runaway Bride} (1999) &
\textit{Rush} (2013) &
\textit{Rush Hour} (1998) \\
\textit{Rush Hour 2} (2001) &
\textit{Saturday Night} (2024) &
\textit{Save the Last Dance} (2001) &
\textit{Saving Mr. Banks} (2013) \\
\textit{Saving Private Ryan} (1998) & 
\textit{Saw} (2003) &
\textit{Saw IV} (2007) &
\textit{Scream 2} (1997) \\
\textit{Scream 3} (2000) & 
\textit{Scream 4} (2011) &
\textit{Se7en} (1995) & 
\textit{Sense and Sensibility} (1995) \\
\textit{Serenity} (2005) & 
\textit{Serial Mom} (1994) &
\textit{Sex and the City} (2008) &
\textit{Shakespeare in Love} (1998) \\
\textit{Shame} (2011) &
\textit{Sherlock Holmes} (2010) &
\textit{Shrek} (2001) & 
\textit{Shrek the Third} (2007) \\
\textit{Sicario} (2015) & 
\textit{Sideways} (2004) &
\textit{Signs} (2002) &
\textit{Silence of the Lambs} (1991) \\
\textit{Silver Linings Playbook} (2012) & 
\textit{Sing Sing} (2024) &
\textit{Single White Female} (2000) &
\textit{Sister Act} (1992) \\
\textit{Six Degrees of Separation} (1993) &
\textit{Sideways} (2004) &
\textit{Skyfall} (2012) & 
\textit{Sleepless in Seattle} (1993) \\ 
\textit{Sleepy Hollow} (1999) &
\textit{Slither} (2006) &
\textit{Slumdog Millionaire} (2008) & 
\textit{Smashed} (2012) \\
\textit{Snatch} (2001) &
\textit{Snow White and the Hunstman} (2012) &
\textit{So I Married an Axe Murderer} (1993) & 
\textit{Solaris} (2002) \\
\textit{Source Code} (2011) &
\textit{South Park} (1999) &
\textit{Spanglish} (2004) &
\textit{Spartan} (2004) \\
\textit{Speed Racer} (2008) &
\textit{Sphere} (1998) &
\textit{Star Trek} (2009) &
\textit{Stepmom} (1998) \\
\textit{Straight Outta Compton} (2015) &
\textit{Strange Days} (1995) &
\textit{Strangers on a Train} (1951) & 
\textit{Sugar} (2009) \\ 
\textit{Sunset Blvd.} (1950) & 
\textit{Sunshine Cleaning} (2009) &
\textit{Superbad} (2007) &
\textit{Supergirl} (1984) \\
\textit{Surrogates} (2009) &
\textit{Sweet Smell of Success} (1957) &
\textit{Swingers} (1996) &
\textit{Swordfish} (2001) \\
\textit{Synecdoche, New York} (2008) &
\textit{Take Shelter} (2011) &
\textit{Taking Lives} (2004) &
\textit{Tamara Drewe} (2010) \\
\textit{Ted} (2012) &
\textit{Tenet} (2020) &
\textit{Terminator} (1984) & 
\textit{Terminator Salvation} (2009) \\ 
\textit{The Addams Family} (1991) & 
\textit{The Adjustment Bureau} (2011) &
\textit{The American} (2010) & 
\textit{The American President} (1995) \\
\textit{The Apartment} (1991) & 
\textit{The Assignment} (1997) &
\textit{The Addams Family} (1991) & 
\textit{The Avengers} (1998) \\
\textit{The Beekeeper} (2024) & 
\textit{The Best Exotic Marigold Hotel} (2012) &
\textit{The Big Lebowski} (1998) & 
\textit{The Big Sick} (2017) \\
\textit{The Birds} (1963)&
\textit{The Blind Side} (2009)&
\textit{The Bling Ring} (2013)&
\textit{The Boondock Saints} (2000)\\
\textit{The Bourne Identity} (2002) &
\textit{The Bourne Ultimatum} (2007) &
\textit{The Box} (2009) &
\textit{The Boxtrolls} (2014) \\ 
\textit{The Breakfast Club} (1985) &
\textit{The Brothers Bloom} (2008) &
\textit{The Butterfly Effect} (2004) &
\textit{The Cell} (2000) \\
\textit{The Cider House Rules} (1999) & 
\textit{The Color Purple} (2023) &
\textit{The Croods} (2013) &
\textit{The Crow} (1994) \\
\textit{The Curious Case of Benjamin Button} (2008) & 
\textit{The Danish Girl} (2016) &
\textit{The Dark Knight Rises} (2012) & 
\textit{The Day the Earth Stood Still} (1951) \\
\textit{The Debt} (2011) & 
\textit{The Deer Hunter} (1978) &
\textit{The Departed} (2006) & 
\textit{The Descendants} (2011) \\
\textit{The Devil Wears Prada} (2006) & 
\textit{The Doors} (1991) &
\textit{The English Patient} (1997) &
\textit{The Family Man} (2000) \\
\textit{The Fault in Our Stars} (2014) &
\textit{The Fifth Element} (1997) & 
\textit{The Fighter} (2010) &
\textit{The Flintstones} (1994) \\
\textit{The French Connection} (1971) &
\textit{The Fugitive} (1993) & 
\textit{The Game} (Not Avail.) &
\textit{The Girl with the Dragon Tattoo} (2011) \\
\textit{The Good Girl} (2002) &
\textit{The Graduate} (1967) &
\textit{The Grapes of Wrath} (1940) &
\textit{The Great Gatsby} (2013) \\
\textit{The Green Mile} (1999) &
\textit{The Grifters} (1990) &
\textit{The Grudge} (2004) &
\textit{The Hangover} (2009) \\
\textit{The Haunting} (1999) &
\textit{The Hebrew Hammer} (2003) &
\textit{The Help} (2011) &
\textit{The Holdovers} (2023) \\
\textit{The Hudsucker Proxy} (1994) &
\textit{The Hunt for Red October} (1990) &
\textit{The Hurt Locker} (2008) &
\textit{The Imaginarium of Doctor Parnassus} (2009) \\
\textit{The Invention of Lying} (2009) &
\textit{The Iron Lady} (2011) &
\textit{The Island} (2005) &
\textit{The Italian Job} (2003) \\
\textit{The Jacket} (1991) &
\textit{The Kids Are All Right} (2010) &
\textit{The Killer} (2023) &
\textit{The King of Comedy} (1982) \\
\textit{The Kingdom} (2007) &
\textit{The Ladykillers} (2004) &
\textit{The Last Boy Scout} (1991) &
\textit{The Last Flight} (1931) \\
\textit{The Last of the Mohicans} (1992) &
\textit{The Last Samurai} (2003) &
\textit{The Last Station} (2009) &
\textit{The LEGO Movie} (2014) \\
\textit{The Lincoln Lawyer} (2011) &
\textit{The Little Mermaid} (1989) &
\textit{The Long Kiss Goodnight} (1996) &
\textit{The Losers} (2010) \\
\textit{The Man Who Knew Too Much} (1956) &
\textit{The Manchurian Candidate} (2004) &
\textit{The Martian} (2015) &
\textit{The Master} (2012) \\
\textit{The Matrix} (1999) &
\textit{The Matrix Reloaded} (2003) &
\textit{The Men Who Stare at Goats} (2009) &
\textit{The Menu} (2022) \\
\textit{The Miracle Worker} (1962) &
\textit{The Mummy} (1999) &
\textit{The Next Three Days} (2010) &
\textit{The Nightmare Before Christmas} (1993) \\
\textit{The Ninth Gate} (1999) &
\textit{The Other Boleyn Girl} (2008) &
\textit{The Pacifier} (2005) &
\textit{The Patriot} (2000) \\
\textit{The Perks of Being a Wallflower} (2012) &
\textit{The Pianist} (2002) &
\textit{The Piano} (1993) &
\textit{The Postman} (1997) \\
\textit{The Prestige} (2006) &
\textit{The Princess Bride} (1987) &
\textit{The Proposal} (2009) &
\textit{The Queen} (2006) \\
\textit{The Reader} (2009) &
\textit{The Replacements} (2000) &
\textit{The Rescuers Down Under} (1990) &
\textit{The Revenant} (2016) \\
\textit{The Rock} (1996) &
\textit{The Rocky Horror Picture Show} (1975) &
\textit{The Roommate} (2011) &
\textit{The Searchers} (1956) \\
\textit{The Secret Life of Walter Mitty} (2013) &
\textit{The Sessions} (2012) &
\textit{The Seventh Seal} (1957) &
\textit{The Shawshank Redemption} (1994) \\
\textit{The Shining} (1980) &
\textit{The Social Network} (2010) &
\textit{The Sting} (2003) &
\textit{The Substance} (2024) \\
\textit{The Talented Mr. Ripley} (1999) &
\textit{The Thing} (1982) &
\textit{The Three Musketeers} (1993) &
\textit{The Time Machine} (2002) \\
\textit{The Tourist} (2010) &
\textit{The Town} (2010) &
\textit{The Truman Show} (1998) &
\textit{The Ugly Truth} (2009) \\
\textit{The Usual Suspects} (1995) &
\textit{The Verdict} (1982) &
\textit{The Visitor} (2008) &
\textit{The Way Back} (2011) \\
\textit{The Whistleblower} (2011) &
\textit{The Wizard of Oz} (1939) &
\textit{The Wolf of Wall Street} (2013) &
\textit{The World is not Enough} (1999) \\
\textit{The Wrestler} (2009) &
\textit{The Zone of Interest} (2023) &
\textit{They} (2002) &
\textit{Thirteen Days} (2000) \\
\textit{Thor} (2011) &
\textit{Three Kings} (1999) &
\textit{Thunderbirds} (2004) &
\textit{Tin Cup} (1996) \\
\textit{Tinker Tailor Soldier Spy} (2011) &
\textit{Titanic} (1997) &
\textit{TMNT} (2007) &
\textit{Tombstone} (1993) \\
\textit{Tomorrow Never Dies} (1997) &
\textit{Top Gun} (1986) &
\textit{Total Recall} (1990) &
\textit{Toy Story} (1995) \\
\textit{Traffic} (2000) &
\textit{Training Day} (2001) &
\textit{Trainspotting} (1996) &
\textit{TRON} (1982) \\
\textit{Tropic Thunder} (2008) &
\textit{True Grit} (2010) &
\textit{True Lies} (1994) &
\textit{True Romance} (1993) \\
\textit{Tucker and Dale vs Evil} (2010) &
\textit{Twilight} (2008) &
\textit{Twin Peaks} (1992) &
\textit{Twins} (1988) \\
\textit{Twister} (1996) &
\textit{Unbreakable} (2000) &
\textit{Under Fire} (1983) &
\textit{Unknown} (2011) \\
\textit{Up} (Not Avail.) &
\textit{Up in the Air} (2009) &
\textit{V for Vendetta} (2006) &
\textit{Valkyrie} (2008) \\
\textit{Vanilla Sky} (2001) &
\textit{Walking Tall} (2004) &
\textit{Wall Street} (1987) &
\textit{Wanted} (2008) \\
\textit{War for the Planet of the Apes} (2017) &
\textit{War Horse} (2011) &
\textit{Warrior} (2011) &
\textit{Watchmen} (2009) \\
\textit{Water for Elephants} (2011) &
\textit{We Own the Night} (2007) &
\textit{What Lies Beneath} (2000) &
\textit{When a Stranger Calls} (1979) \\
\textit{While She Was Out} (2008) &
\textit{Whiplash} (2014) &
\textit{White Christmas} (1954) &
\textit{Whiteout} (2009) \\
\textit{Wicked} (2024) &
\textit{Wild At Heart} (1990) &
\textit{Wild Hogs} (2007) &
\textit{Wild Things} (1998) \\
\textit{Wild Wild West} (1999) &
\textit{Willow} (1988) &
\textit{Win Win} (2011) &
\textit{Witness} (1985) \\
\textit{Wonder Boys} (2000) &
\textit{Wonka} (2023) &
\textit{xXx} (2002) &
\textit{Year One} (2009) \\
\textit{Yes Man} (2008) &
\textit{Youth in Revolt} (2010) &
\textit{Zero Dark Thirty} (2013) &
\textit{Zootopia} (2016) \\
\end{longtable}
}
\end{center}

\end{document}